\documentclass[10pt]{article}
\usepackage{graphicx}
\usepackage{amssymb}
\usepackage{epstopdf}
\usepackage{fancyhdr}
\usepackage{amsmath}
\usepackage{amsthm}
\usepackage{amsfonts}
\usepackage{enumerate}
\usepackage{latexsym}
\usepackage{float}
\usepackage{color}
\usepackage{bm}
\usepackage{ifthen}

\usepackage[pdfencoding=auto, psdextra]{hyperref}
\hypersetup{
    colorlinks=false, 
    linktoc=all,     
    linkcolor=blue,  
}

\def\beq{\begin{equation}}  
\def\eeq{\end{equation}}  

\def\Tr{\text{Tr}}
\def\d{\partial}

\begin{document}

\title{Glueball-Meson Mixing in Holographic QCD}

\author{Sophia K.~Domokos\footnote{sdomokos@nyit.edu} \ and Nelia Mann\footnote{mannn@union.edu}}
\date{%
$^1$ Department of Physics, New York Institute of Technology,\\ 16 W. 61st Street, New York, NY 10023\\
$^2$ Department of Physics and Astronomy, Union College, Schenectady, NY 12308 }

\maketitle

\begin{abstract}
Top-down holographic QCD models often work in the ``probe" (or ``quenched") limit, which assumes that the number of colors is much greater than the number of flavors. Relaxing this limit is essential to a fuller understanding of holography and more accurate phenomenological predictions. In this work, we focus on a mixing of glueball and meson mass eigenstates that arises from the DBI action as a finite $N_f/N_c$ effect. For concreteness, we work in the Witten-Sakai-Sugimoto model, and show that this mixing must be treated in conjunction with the backreaction of the flavor branes onto the background geometry.  Including the backreaction with the simplification that it is ``smeared out'' over the compact transverse direction, we derive a corrected effective action for the vector glueball and scalar states. Along the way, we observe a St\"{u}ckelberg-like mechanism that restores translation invariance in the transverse direction. We also derive a general technique, that lends itself easily to numerics, for finding mass eigenstates of Lagrangians with vector-scalar mixing. We then calculate the first order corrections to the mass spectra of both the vector and scalar particles, and show that the term that explicitly mixes vector and scalar states is the most significant correction to the masses of low-lying scalar mesons.
\end{abstract}

\newpage

\tableofcontents

\section{Introduction}

The past decades have seen many advances in moving holographic QCD beyond the strict  $N_f\ll N_c$ limit, which corresponds to the quenched approximation on the lattice, where fermions are not allowed to run in loops. In holographic QCD, this is equivalent to neglecting the effect of the flavor degrees of freedom on the supergravity background. For top-down constructions 
built from intersections of ``color" and ``flavor" branes like the Witten-Sakai-Sugimoto model (WSS) \cite{Witten:1998zw, Sakai2005}, it means not only neglecting the backreaction of the flavor branes in the background generated by the color branes, but also suppressing interactions between mesons (brane fields) and glueballs (bulk fields). 

Given that $N_f\sim N_c$ in real QCD, moving beyond the strict $N_f\ll N_c$ limit not only produces more precise predictions for hadron physics, it allows holographic models to make predictions for processes that allow (indirect) observation of glueballs via decay to mesons (see e.g. \cite{Hashimoto:2007ze, Brunner:2015oqa}), and, as is our focus here,  mixing between meson and glueball mass eigenstates. In addition, one might hope that finite $N_f/N_c$ effects will
ameliorate some of the problems plaguing (especially top-down) holographic QCD, such as the large number of light, spurious states in the glueball \cite{BROWER2000} and meson \cite{Sakai2005} spectra. From the phenomenological side, it has long been conjectured that the light isospin 0 mesons mix with glueball states \cite{Amsler:1995td, Close:2001ga, Janowski:2014ppa, ParticleDataGroup:2020ssz}.  

Finding exact supergravity solutions that do include flavor backreaction remains a significant challenge, because brane intersections boast fewer isometries than single stacks of branes and thus require substantially more complicated solutions.
Some progress has been made in the Veneziano limit, where $N_f/N_c$ is fixed and finite, and $N_f,~N_c\rightarrow\infty$.  Closed-form solutions with spatially localized flavor branes have been found, but only in supersymmetric models such as \cite{Burrington:2004id, Kirsch:2005uy, Jokela:2021evo}. Most works eliminate the difficulty of reduced isometry by artificially enlarging the isometry group, that is, by ``smearing" the flavor branes transverse to their worldvolumes. (See for instance  \cite{Benini:2006hh} with smeared D7-flavor branes in the Klebanov-Witten model, smeared D6-flavors in ABJM \cite{Bea:2013jxa}, and \cite{Nunez:2010sf} for a beautiful review of smeared Veneziano-limit backgrounds.) An alternative option, that has yielded many interesting results over the past several years, is to take the Veneziano limit  \cite{Gursoy:2010fj} in a simpler, bottom-up model: in Improved Holographic QCD (IHQCD)  \cite{Gursoy:2010fj}. 

Incorporating flavor physics in  Witten-Sakai-Sugimoto (WSS) is much more difficult. It has, however, been attempted at leading order in $N_f/N_c$ -- with Burrington, Sonnenschein and Kaplunovsky \cite{Burrington2008} tackling localized flavor branes, and Bigazzi and Cotrone \cite{Bigazzi2015} smeared ones. These backgrounds are the basis of our approach.

In this paper, we study an example of glueball-meson mixing in the WSS model \cite{Witten:1998zw, Sakai2005}.
 We focus on mixing between a vector glueball, dual to a mode of the bulk graviton, and a (pseudo)scalar meson, dual to transverse flavor-brane fluctuations. The effect arises due to a quadratic term in the DBI action that couples graviton modes with brane modes, as pointed out in \cite{Domokos2017}.  We work in the ``smeared" limit of \cite{Bigazzi2015}, where the brane scalar and bulk vector modes decouple from all other excitations at quadratic order. 
 
 We will show not only that the mixing effect arises at the same order in $N_f/N_c$ as the first-order backreaction of the flavor-branes on the geometry, but also that including both effects is necessary  to generate a physically sensible Lagrangian, and to preserve the translation symmetry broken by the probe branes and restored in the smeared approximation.  We also establish a general method for finding the mass eigenvalues for a Lagrangian with vector-scalar mixing, in which the result involves scalar fields that have kinetic and mass terms that cannot be simultaneously diagonalized. We show that the mass spectrum of the  vector glueball is unaffected by the mixing term, while the scalar mass spectrum may be altered substantially.
 
 Though the states we study appear to be spurious from the perspective of the lattice and glueball spectra, our work serves not only as a ``warm-up" for a comprehensive treatment of backreaction on the hadron spectrum, it is also relevant to other holographic models relying on brane intersections, like the famous D3-D5 model\cite{Karch:2000gx} often used in AdS/CMT.

We should note that some other examples of glueball-meson mixing have been explored by other authors.  Rinaldi et al. used a bottom-up model to argue that 
glueball and meson states above 2 GeV experience very little mixing \cite{Rinaldi2020PureGS}. Leutgleb and Rebhan, meanwhile, pointed to a glueball-meson mixing phenomenon from the flavor-brane Chern-Simons term in WSS, which involved a glueball dual to a bulk Ramond-Ramond 1-form, and the $\eta$ meson (dual to a mode of the brane gauge field) \cite{Rebhan2020}.

The outline of this paper is as follows: In Section \ref{sec:background} we  briefly review the original WSS model, then describe the backreacted geometries of \cite{Burrington2008, Bigazzi2015}. In Section \ref{sec:excitations} we derive the mixed quadratic-order action for the vector glueball and brane scalar, identifying the realization of a residual translation symmetry. We establish a method for finding the mass eigenvalues of our glueball and meson states in Section \ref{sec:KK}, and identify the linear order corrected masses of the corresponding hadrons, discussing the trends we find there in Section \ref{sec:results}. In Section \ref{sec:conclusion} we conclude and describe directions for future work. We relegate tedious but important details to a series of Appendices. 

\section{\label{sec:background} Review of the WSS Model}

We first give an brief overview of the WSS model without and with the leading-order backreaction, before turning to technical details in the next subsections. We also take this  opportunity to establish our variables and conventions. 

\subsection{Overview of the WSS Model}
The WSS model is one of the most commonly used holographic QCD frameworks, due to its elegant and intuitive geometric realization of confinement \cite{Witten:1998zw} and chiral symmetry-breaking \cite{Sakai2005}. The model is based on a non-supersymmetric brane intersection, in which $N_c$ D4-branes provide the color SU($N_c$), while parallel stacks of $N_f$ D8- and $\overline{\text{D8}}$-branes provide the chiral U($N_f)_L\times$ U($N_f)_R$ flavor group. Here (as in \cite{Sakai2005}) we focus on the case where the flavor branes are coincident, and the quark mass is zero. 

Both stacks of branes are extended in the (3+1) dimensions of the dual field theory, and are orthogonal on the remaining 6 dimensions as shown in Table \ref{tab:SSbranes}. In particular, the D4-branes are extended along $\tau$, taken to be periodic as $\tau\sim\tau+\delta\tau$. 
\begin{table}[h!]
\centering
\begin{tabular}{c |  c c c | c | c | c c c c|} 
  &  & $x^i$ & & $\tau$ & $U$ & & $x^\alpha$ & \\
  \hline
 D4 & X & X &  X & X & & & & &  \\ 
 D8, $\overline{\text{D8}}$ & X & X & X & & X & X & X & X & X\\
 \hline
 \end{tabular}
\caption{Brane configuration and coordinate labels in the WSS model. $x^\mu = (x^0,x^i)$ denotes the coordinates common to both color and flavor branes, which span the field theory directions. The $N_c$ color D4-branes wrap the compact $\tau$ direction, where the $N_f$ flavor D8's are localized in this direction. The flavor branes are extended in the remaining 5 directions: a radial coordinate $U$ and an $S^4$ parameterized by $x^\alpha$.}\label{tab:SSbranes}
\end{table}

In the  $N_c\gg 1$ and $N_c\gg N_f$ limit, the D4-branes curve the spacetime, generating a confining supergravity background. As in \cite{Sakai2005}, antisymmetric boundary conditions imposed on the fermions living on the flavor branes along the compact $\tau$ direction lift the masses of these states so they do not appear in the light spectrum. 
 
When the flavor branes are treated as probes, the $\text{D8}$ and $\overline{\text{D8}}$-branes assume a non-trivial profile in this background, joining deep in the space (in the IR of the field theory) but remaining parallel and separated near the boundary (in the UV). This realizes the breaking of U($N_f)_L\times$ U($N_f)_R\rightarrow$ U($N_f)_V$ at low energies. The closed string degrees of freedom (or, at low energies, the supergravity modes) correspond to glueball states. Open string degrees of freedom (or the brane fields) correspond to mesons.

Both Burrington et al.\cite{Burrington2008} and Bigazzi \& Cotrone\cite{Bigazzi2015} furnish first order in $N_f/N_c$ corrections to the supergravity background generated by the D4-branes.  In the  comprehensive work of \cite{Burrington2008}, the flavor branes are localized in the $\tau$ direction, and  solutions are given as a Fourier decomposition around the $\tau$ circle. In \cite{Bigazzi2015},  the flavor branes are smeared along the $\tau$ circle, maintaining the same isometry as the original, un-backreacted geometry, which permits them to identify analytic solutions.  

The physical interpretation of smearing the flavor branes around the $\tau$ circle is somewhat mysterious, especially in the WSS context, where it implies coincident branes and antibranes. We believe that the proper interpretation of smearing and of \cite{Bigazzi2015}'s result is actually as the zeroth $\tau$-direction Fourier mode of the full, $\tau$-dependent solution from \cite{Burrington2008}. Indeed, one can check that truncating \cite{Burrington2008}'s ansatz on the constant mode in the $\tau$ direction yields the same equations of motion as \cite{Bigazzi2015}'s. (As discussed below, the background solutions of the two papers differ, however, due to diverging choices of boundary conditions.)

The spectrum of excitations above this background contains Kaluza-Klein (KK) towers on  the $\tau$ circle, which for $\tau$-dependent backgrounds should yield highly non-trivial mixing  among many different levels in the KK towers.  However, we will see that restricting to the trivial mode of the background in the $\tau$ direction also allows us to truncate the graviton excitations to the lowest KK mode, as well as limiting the components of the graviton we need to consider. This radically simplifies our analysis.  Specifically, it allows us to isolate a glueball-meson mixing between just two fields.  Because of this simplification, in this work we use the constant $\tau$-direction Fourier mode of the background in \cite{Burrington2008} (or equivalently, \cite{Bigazzi2015}). 

We now provide some further details and establish conventions for the WSS model with and without backreaction.

\subsection{D4-brane Background}

The supergravity background generated by the $N_c$ D4-branes is determined by the supergravity action given in string frame as
\beq
\label{eqn:Sbulk}
S_{\mathrm{bulk}} = \frac{1}{2\kappa_{10}^2}\int d^{10}x \sqrt{-\det g}\left[e^{-2\phi}\left(R - \frac{1}{12}H_{MNL}H^{MNL} + 4\partial_{M}\phi\partial^{M}\phi\right) - \frac{1}{2\cdot 4!}F_4^2\right] \, ,
\eeq
where $g_{MN}$ is the metric, $R$ is the Ricci scalar, $F_4$ is the 4-form sourced by the D4-branes, and $\phi$ is the dilaton. $H_{MNL}$ is the field strength for the Kalb-Ramond field $B_{MN}$, which we ignore in what follows: it vanishes on the D4-brane background and on the smeared, first-order-backreacted background described in the next subsection. It also does not couple at quadratic order to the excitations we are interested in. 
The indices $N$ and $M$ run over the 10 spacetime coordinates, and the Newton constant is given by $\kappa_{10}^2 = \frac{(2\pi\ell_s)^8}{4\pi}$, where $\ell_s$ is the string length.

The  equation of motion for the metric,
\beq
0 = \tilde{\mathcal{G}}^{(0)}_{MN} = \tilde{R}^{(0)}_{MN} - \tilde{g}^{(0)}_{MN}\left[\frac{1}{2}\tilde{R}^{(0)} + 2(\tilde{\nabla}^{(0)})^{2}\tilde{\phi}^{(0)} -  2\Big(\tilde{\nabla}^{(0)}\tilde{\phi}^{(0)}\Big)^2 - \frac{1}{4\cdot 4!}e^{2\tilde{\phi}^{(0)}}\Big(\tilde{F}_4^{(0)}\Big)^2\right] + 2\tilde{\nabla}^{(0)}_M\tilde{\nabla}^{(0)}_{N}\tilde{\phi}^{(0)}~,
\eeq
will come in handy below. Here we are defining $\tilde{\mathcal{G}}^{(0)}$ as the ``supergravity version'' of the Einstein tensor -- that is, as the combination of Riemann tensor, dilaton, and 4-form field strength appearing on the right-hand side. This combination vanishes on the D4-brane background.

Here we are introducing the notation $g_{MN}$ for the metric including graviton modes, $\tilde{g}_{MN}$ for the background metric, and $\tilde{g}_{MN}^{(0)}$ for the unbackreacted background metric -- that is, the probe limit. We use the same conventions for the Ramond-Ramond forms and the dilaton.

The near-horizon geometry of the D4-branes satisfies this equation as well as equations for the dilaton and Ramond-Ramond forms. The solution is given by
\begin{align}\label{SSstrframe}
&ds_{(0)}^2:= \tilde{g}^{(0)}_{MN}dx^M dx^N=\left(\frac{U}{R} \right)^{3/2}\left( \eta_{\mu\nu}dx^\mu dx^\nu + f(U)d\tau^2\right)+\left(\frac{U}{R} \right)^{-3/2}\left(\frac{dU^2}{f(U)}+U^2d\Omega_4^2 \right) \cr
&e^{\tilde{\phi}^{(0)}}=g_s\left(\frac{U}{R} \right)^{3/4}~,\quad \tilde{F}^{(0)}_4=d\tilde{C}^{(0)}_3 =\frac{2\pi N_c}{V_4}\epsilon_4 ~, \quad f(U) = 1-\frac{U_{KK}^3}{U^3} \, , 
\end{align}
with $\epsilon_4$ the volume form on the 4-sphere, and $R^3=\pi g_sN_c\ell_s^3$. The volume of the 4-sphere is
$V_4 = \frac{8\pi^2}{3}.$

Recall that $\tau$ is periodic, with $\tau\sim \tau+\delta\tau$. The radial coordinate $U$ transverse to the D4 stack ranges over   $U\in [U_{\textrm{KK}},\infty)$, where $U_{\textrm{KK}}>0$.  In order to avoid a conical singularity we must have $\delta\tau =\frac{4\pi}{3}\frac{R^{3/2}}{U_{KK}^{1/2}}$.  We often also use the KK scale in the periodic $\tau$ direction, $M_{\textrm{KK}} = 2\pi/\delta\tau$. 

In terms of field theory quantities, the constants appearing in the background are
\begin{align}
    R^3=\frac{g_{YM}^2N_c\ell_s^2}{M_{KK}}~,\quad U_{\mathrm{KK}}=\frac{2}{9}g_{YM}^2 N_c M_{\textrm{KK}}\ell_s^2~,\quad g_s=\frac{g_{YM}^2}{2\pi M_{\textrm{KK}}\ell_s}~, 
\end{align}
where $g_{YM}$ is the effectively 4d Yang-Mills coupling of the field theory (below the $M_{\mathrm{KK}}$ scale). The supergravity description is reliable while 
$1\ll N_cg_{YM}^2\ll 1/g_{YM}^4$ \cite{Witten:1998zw, Sakai2005}.

\subsection{Adding Probe Flavor Branes}

Adding stacks of $N_f$ D8- and $N_f$ $\overline{\text{D8}}$-branes to the D4 background above realizes QCD's SU($N_f)_L\times$SU($N_f)_R$ flavor symmetry. The D8 and  $\overline{\textrm{D8}}$-branes are separated along the $\tau$ direction and  extended along radial coordinate $U$, along the ``field theory directions'' $x^{\mu}$, and along the $S^4$.  In the probe limit, the brane and anti-brane stacks find an  energy-minimizing configuration  by joining at a finite radial value $U=U_0$, thus tracing out a curve in the $U$ and $\tau$ directions (representing the breaking of chiral symmetry) \cite{Witten:1998zw, Sakai2005}. 

We choose to embed the branes with the ``maximal" embedding, such that $U_0 = U_{\mathrm{KK}}$. The embedding function thus simplifies so that the D8 and $\overline{\textrm{D8}}$ branes are at antipodal points on the $\tau$ circle.  This makes the $\tau$ direction transverse to the branes.
Especially when discussing the symmetry properties of brane modes, it will sometimes be helpful to use the coordinate $Z \in [-\infty, \infty]$, related to $U$ as
\beq
\frac{U}{U_{\mathrm{KK}}} = \left(1 + \frac{Z^2}{U_{\mathrm{KK}}^2}\right)^{1/3} \, . 
\eeq
The $Z$-coordinate is natural on the branes, where the (anti)symmetry of the normalizeable modes determines the parity and charge-conjugation quantum numbers of the corresponding mesons, as detailed in \cite{Sakai2005}. However, $Z$ double-covers the $U$-coordinate, which is more natural in the bulk. 

The degrees of freedom on the D8-branes consist of a non-abelian U($N_f$) gauge field and a single, U($N_f$)-valued scalar $\Phi$, representing transverse fluctuations of the flavor branes in the $\tau$ direction.

At low energies, the physics on the branes is described by the Dirac-Born-Infeld (DBI) and Chern-Simons (CS) actions. Note that these actions encode not only terms involving the brane degrees of freedom, they also includes couplings between the bulk graviton, dilaton, and Ramond-Ramond forms and the brane fields. 

The Chern-Simons term plays no role in the present work, as argued in Appendix \ref{sec:CSappendix}. The DBI action, meanwhile, is given by
\beq
S_{\mathrm{DBI}} = -\frac{2\pi}{(2\pi\ell_s)^9}\int d^9x \, \mathrm{Tr}\left[e^{-\phi}\sqrt{-\det\left(P[g_{ab}] + 2\pi\ell_s^2 F_{ab}\right)}\right]~,
\eeq
where $F_{ab}$ is the field strength for the gauge field on the brane, and $P[g_{ab}]$ is the pullback of the metric onto the brane.
The indices $\{a, b\}$ run over the coordinates along the branes.  

 We work in static gauge and in the extremal configuration of the flavor brane stack, so the pullback of background metric is trivial for directions along the brane, but has nontrivial coontributions from the scalar $\Phi$ encoding transverse fluctuations in the $\tau$ direction.  It is given by
\beq
\label{eqn:pullback}
P[g_{ab}] = g_{ab} + \sqrt{2\pi\ell_s^2} g_{\tau a}\partial_b\Phi + \sqrt{2\pi\ell_s^2} g_{\tau b}\partial_a\Phi + 2\pi\ell_s^2 g_{\tau\tau}\partial_a\Phi \partial_b\Phi \, , 
\eeq
where we have dropped terms associated with the bulk Kalb-Ramond $B$-field and the brane gauge field, as they are not relevant here. Our convention fixes the units of $\Phi$ to be inverse-mass (as for the gauge field).

 When one expands the DBI action order by order in field fluctations, the leading terms are in fact {\em linear} (tadpole) terms in the dilaton and graviton. This is not surprising: after all, the D8 branes act as sources for the dilaton and graviton, and the current supergravity background is a solution to the Einstein equations that only takes into account the masses of the D4-branes (as noted in \cite{Domokos2017}). In the next subsection, we will see that these tadpoles indeed disappear when one takes into account the backreaction of the flavor branes.

From this point on, we will ignore terms in the brane action associated with the gauge field, since it does not couple at quadratic order to the excitations of interest.  We will also restrict focus on the $U(1)$ sector, so $\Phi$ denotes just the U(1) part of the scalar field, and the trace over flavor indices in the DBI action simply contributes an overall factor of $N_f$.

The parts of the DBI action relevant to our story thus become
\beq
\label{eqn:SDBI}
S_{\mathrm{DBI}} \supset-\frac{\tilde{\zeta}\delta\tau}{2\kappa_{10}^2}\int d^9x \, e^{-\phi}\sqrt{-\det P[g_{ab}]} \, ,
\eeq
where we define parameters $\tilde{\zeta}$, and $\zeta$ as
\beq
\tilde{\zeta}  = \frac{9}{8g_s U_{\mathrm{KK}}^2}\left(\frac{U_{\mathrm{KK}}}{R}\right)^{3/2} \frac{(N_cg_{\mathrm{YM}}^2)^2N_f}{27\pi^3N_c} :=  \frac{9\zeta}{8g_s U_{\mathrm{KK}}^2}\left(\frac{U_{\mathrm{KK}}}{R}\right)^{3/2}~,
\eeq
and
\beq
\zeta = \frac{(N_cg_{\mathrm{YM}}^2)^2}{27\pi^3}\left(\frac{N_f}{N_c}\right)~.
\eeq
We will see that $\zeta$ (or equivalently $\tilde{\zeta}$) controls both the backreaction of the flavor branes on the background and the strength of the mixing between glueball and meson modes. 

To get an idea of the rough size of $\zeta$ in QCD, note that the low-energy limit of the WSS model relies on two free parameters: $M_{\mathrm{KK}}$ and the effective 4D 't Hooft coupling $N_c g_{\mathrm{YM}}^2$.  
The WSS model's predictions for the pion decay constant and the $\rho$ mass, 
    \beq
    f_{\pi}^2 = \Big(.318\Big)\frac{g_{\mathrm{YM}}^2N_c^2M_{\mathrm{KK}}^2}{54\pi^3}
    \eeq
    and 
    \beq
    m_{\rho} = \Big(.817\Big)M_{\mathrm{KK}} \, ,
    \eeq
are often used to fix $M_{\mathrm{KK}}$ and $N_c g_{\mathrm{YM}}^2$. 
With the $N_f = N_c = 3$, this gives $\zeta = 0.33$. 



\subsection{Backreaction of the D8-Branes}

We now move beyond the original WSS model to include the leading order backreaction of the flavor D8-branes by solving the equations of motion derived from  $S_{\mathrm{bulk}} + S_{\mathrm{DBI}}$, as in  \cite{Burrington:2004id, Bigazzi2015}.

We denote the backreacted versions of the background metric, dilaton, and Ramond-Ramond four-form field strength as $\tilde{g}_{MN}$, $\tilde{\phi}$, and $\tilde{F}_4$, respectively.  
The new Einstein equation is given by perturbing the metric as $g_{MN}=\tilde{g}_{MN}+\delta g_{MN}$ and expanding to leading order in $\delta g_{MN}$:
\begin{align}
\delta \bigg( S_{\mathrm{bulk}} + S_{\mathrm{DBI}} \bigg) &= -\frac{1}{2\kappa_{10}^2}\int d^{10}x \sqrt{-\det\tilde{g}} \, e^{-2\tilde{\phi}} \, \tilde{\mathcal{G}}_{MN}\, \delta g^{MN} \cr
&\qquad\qquad\qquad -\frac{\tilde{\zeta}\delta\tau}{4\kappa_{10}^2}\int d^{10}x \, \Big(\delta(\tau) + \delta(\tau + \delta\tau/2)\Big) \frac{e^{-\tilde{\phi}}\sqrt{-\det\tilde{g}}}{\sqrt{\tilde{g}_{\tau\tau}}}\left(\tilde{g}_{ab} \, \delta g^{ab}\right)~.
\end{align}
where as before $\tilde{\mathcal{G}}_{MN}$ is the ``supergravity version'' of the background Einstein tensor and we have rewritten the DBI action as an integral over the bulk coordinates. The modified equation of motion for $\tilde{g}_{ab}$ is thus
\beq
\tilde{\mathcal{G}}_{ab} = -\frac{\tilde{\zeta}\delta\tau}{2}\Big(\delta(\tau) + \delta(\tau + \delta\tau/2)\Big)\frac{e^{\tilde{\phi}}}{\sqrt{\tilde{g}_{\tau\tau}}} \, \tilde{g}_{ab} \, .
\eeq
$\tilde{\phi}$ is also modified. The source term in the Einstein equation corresponds to a tadpole term for the graviton on the D8-brane worldvolume. Solving the new Einstein equations with this source term included represents a leading-order backreacted solution, and will eliminate the tadpole terms as a result.

The source term representing the flavor branes is not uniform along the $\tau$ direction. Indeed, the leading order $\tau$-varying backreaction was worked out in \cite{Burrington2008} as a Fourier mode expansion along the $\tau$ direction.  As noted previously,  we consider only the trivial $\tau$-direction Fourier mode in what follows -- equivalent to the ``smeared" approximation. This allows us to consider a decoupled sector of the excitations.

In the smeared approximation, the equation of motion for $\tilde{g}_{ab}$ becomes simply
\beq
\label{eqn:BREOM}
\tilde{\mathcal{G}}_{ab} = -\tilde{\zeta}\frac{e^{\tilde{\phi}}}{\sqrt{\tilde{g}_{\tau\tau}}} \, \tilde{g}_{ab} \, .
\eeq

We can also assume from now on that the backreacted background metric, dilaton, and potential depend only on the radial coordinate $U$ (and, in particular, are $\tau$-independent), and also that the background metric remains diagonal. In other words, we assume that the backreacted background obeys the same isometries as the original, un-backreacted version, with the backreaction just modifying the functions of $U$ appearing in equation \eqref{SSstrframe}. Because the D8-branes do not directly source a $C_3$, the field strength $\tilde{F}_4 = \tilde{F}_4^{(0)}$ remains unchanged. 

It might appear at this point that equation \eqref{eqn:BREOM} is ``exact in $\zeta$'', and thus that no approximation to leading order in the backreaction has been made.  However, one should remember that in the process of arriving at this equation, we began with just the D4-branes, found the geometry they source, and then took a near-horizon limit to obtain the supergravity action \eqref{eqn:Sbulk}, before adding in the D8-branes.  To obtain equations truly exact in $\zeta$, we would have to begin by treating the D4 and D8 branes on an equal footing, find the geometry sourced by both, and then take a near-horizon limit of that.  As a result, we should interpret \eqref{eqn:BREOM} as already assuming that $\zeta$ is small.

The specific results for the background to linear order in $\zeta$, as derived by \cite{Bigazzi2015}, are summarized in Appendix \ref{sec:LinearBackreactionApp}.

\section{Excitations Around the Background}\label{sec:excitations}

We now turn to the a subset of the fluctuations (mesons and glueballs) on this backreacted background: the brane scalar $\Phi$, and gravitons $h_{MN}$ defined as
\beq
\label{eqn:graviton}
g_{MN}  = \tilde{g}_{MN} + h_{MN} \, .
\eeq
 We focus in particular on a piece of the 10D graviton $\{h^{\tau}_{\mu}, h^{\tau}_{U}\}$ which transforms as a 5D vector in $(x^\mu, U)$.

The graviton should be expanded in terms of Kaluza-Klein (KK) modes along the $\tau$ and $S^4$ directions, which indeed correspond to higher mass and spin glueball  states. We can, however, neglect all but the trivial KK modes of the graviton in what follows provided that the background as no $\tau$ dependence -- as is the case in the smeared approximation we use. (If one allowed the background to depend on $\tau$, the different KK modes would mix with each other.)  Similarly, we can restrict to the zeroeth mode in the KK tower associated on the $S^4$.

We thus will move forward assuming each excitation field is a function only of the Lorentz coordinates $x^{\mu}$, and the $U$ (or equivalently $Z$) coordinate.

\subsection{Expanding the Bulk Action}
 We now derive an effective  Lagrangian for these modes by expanding the bulk and DBI actions to quadratic order in fields.
Expanding the bulk action, we have
\beq
\label{eqn:d2Sbulk}
S_{\mathrm{bulk}} = -\frac{1}{2\kappa_{10}^2}\int d^{10}x \, \sqrt{-\det \tilde{g}} \, e^{-2\tilde{\phi}} \, \tilde{g}_{\tau\tau} \times
\eeq
$$
\Bigg\{\frac{1}{4}\tilde{g}^{\mu\rho}\tilde{g}^{\nu\sigma}\Big(\partial_{\rho}h^{\tau}_{\sigma} - \partial_{\sigma}h^{\tau}_{\rho}\Big)\Big(\partial_{\mu}h^{\tau}_{\nu} - \partial_{\nu}h^{\tau}_{\mu}\Big)
 +\frac{1}{2}\tilde{g}^{\mu\nu}\tilde{g}^{UU}\Big(\partial_{\mu}h^{\tau}_{U} - \partial_{U}h^{\tau}_{\mu}\Big)\Big(\partial_{\nu}h^{\tau}_{U} - \partial_{U}h^{\tau}_{\nu}\Big)
+ \tilde{\mathcal{G}}^{UU}h^{\tau}_{U}h^{\tau}_{U} + \tilde{\mathcal{G}}^{\mu\nu}h^{\tau}_{\mu}h^{\tau}_{\nu}\Bigg\} \, .
$$
(See Appendix \ref{sec:BulkActionExpn} for the details.) Indices are raised and lowered using the background metric, $\tilde{g}_{MN}$. Note the appearance of the background supergravity ``Einstein tensor" $\tilde{\mathcal{G}}$, which allows to plug in the equation of motion \eqref{eqn:BREOM} satisfied by the backreacted background. This gives
\beq
S_{\mathrm{bulk}} = \frac{1}{2\kappa_{10}^2}\int d^{10}x \, \sqrt{-\det \tilde{g}} \, e^{-2\tilde{\phi}} \, \tilde{g}_{\tau\tau} \, 
\Bigg\{-\frac{1}{4}\tilde{g}^{\mu\rho}\tilde{g}^{\nu\sigma}\Big(\partial_{\rho}h^{\tau}_{\sigma} - \partial_{\sigma}h^{\tau}_{\rho}\Big)\Big(\partial_{\mu}h^{\tau}_{\nu} - \partial_{\nu}h^{\tau}_{\mu}\Big)
\eeq
$$
 -\frac{1}{2}\tilde{g}^{\mu\nu}\tilde{g}^{UU}\Big(\partial_{\mu}h^{\tau}_{U} - \partial_{U}h^{\tau}_{\mu}\Big)\Big(\partial_{\nu}h^{\tau}_{U} - \partial_{U}h^{\tau}_{\nu}\Big)
- \frac{\tilde{\zeta}e^{\tilde{\phi}}}{\sqrt{\tilde{g}_{\tau\tau}}}\tilde{g}^{UU}h^{\tau}_{U}h^{\tau}_{U} - \frac{\tilde{\zeta}e^{\tilde{\phi}}}{\sqrt{\tilde{g}_{\tau\tau}}}\tilde{g}^{\mu\nu}h^{\tau}_{\mu}h^{\tau}_{\nu}\Bigg\} \, ,
$$
making it clear that mass terms for $\{h^{\tau}_{\mu}, h^{\tau}_{U}\}$ appear as leading order corrections due to the back reaction.

Next, we switch to a dimensionless radial coordinate, with which 
\beq
u = \frac{U}{U_{\mathrm{KK}}} \, , \hspace{.5in} h^{\tau}_{u} = U_{\mathrm{KK}}h^{\tau}_{U} \, , \hspace{.5in} f(u) = 1 - u^{-3} \, ,
\eeq
and for convenience rewrite the action in terms of three  functions $\{a(u), b(u), c(u)\}$ defined as 

\beq
\frac{g_s^2}{U_{\mathrm{KK}}R^3}\sqrt{-\frac{\det \tilde{g}}{\det s}} \, e^{-2\tilde{\phi}} \, \tilde{g}_{\tau\tau}\tilde{g}^{\mu\rho}\tilde{g}^{\nu\sigma} = a(u) \, \eta^{\mu\rho}\eta^{\nu\sigma} \, ,
\eeq
\beq
\frac{g_s^2}{U_{\mathrm{KK}}R^3}\sqrt{-\frac{\det \tilde{g}}{\det s}} \, e^{-2\tilde{\phi}} \, \tilde{g}_{\tau\tau}\frac{\tilde{g}^{\mu\nu}\tilde{g}^{UU}}{U_{\mathrm{KK}}^2} = M_{\mathrm{KK}}^2 a(u)b(u)\eta^{\mu\nu} \, , 
\eeq
and 
\beq
\frac{g_s^2}{U_{\mathrm{KK}}R^3}\sqrt{-\frac{\det\tilde{g}}{\det s}} \, e^{-2\tilde{\phi}} \, \tilde{g}_{\tau\tau} \, \frac{\tilde{\zeta}e^{\tilde{\phi}}}{\sqrt{\tilde{g}_{\tau\tau}}} \, \tilde{g}^{\mu\nu} = \frac{\zeta M_{\mathrm{KK}}^2}{2} \, c(u) \,  \eta^{\mu\nu} \, ,
\eeq
where $\det s$ is the determinant of the metric on the four-sphere.  These expressions are essentially equivalent to defining the functions in $\tilde{g}_{\tau\tau}$, $\tilde{g}_{\mu\nu}$, and $\tilde{g}_{UU}$.  On the original un-backreacted WSS background, 
\beq
a^{(0)}(u) = uf(u) \, , \hspace{.5in} b^{(0)}(u) = \frac{4}{9}u^3f(u) \, , \hspace{.5in} c^{(0)}(u) = u^{5/2}\sqrt{f(u)} \, .
\eeq
Armed with these definitions, we integrate out over the $\tau$ direction as well as the 4-sphere, yielding an effectively 5D bulk action,
\beq
\label{eqn:Sbulkclean}
S_{\mathrm{bulk}} = K\int d^{4}x \, du \, 
 \Bigg\{-\frac{a(u)}{4}\eta^{\mu\rho}\eta^{\nu\sigma}\Big(\partial_{\rho}h^{\tau}_{\sigma} - \partial_{\sigma}h^{\tau}_{\rho}\Big)\Big(\partial_{\mu}h^{\tau}_{\nu} - \partial_{\nu}h^{\tau}_{\mu}\Big)
 \eeq
 $$
 -\frac{M_{\mathrm{KK}}^2a(u)b(u)\eta^{\mu\nu}}{2}\Big(\partial_{\mu}h^{\tau}_{u} - \partial_{u}h^{\tau}_{\mu}\Big)\Big(\partial_{\nu}h^{\tau}_{u} - \partial_{u}h^{\tau}_{\nu}\Big)
 - \frac{M_{\mathrm{KK}}^2\zeta c(u)}{2}\eta^{\mu\nu}h^{\tau}_{\mu}h^{\tau}_{\nu} - \frac{M_{\mathrm{KK}}^4\zeta b(u)c(u)}{2}h^{\tau}_{u}h^{\tau}_{u}\Bigg\} \, , 
$$
where the overall constant $K$ is defined as
\beq
K = \frac{V_4\delta\tau \, U_{\mathrm{KK}}^2R^3}{2\kappa_{10}^2g_s^2} \, .
\eeq
This form makes clear that  $\{h^{\tau}_{\mu}, h^{\tau}_{u}\}$ indeed transforms as a 5D vector on a warped background (expressed through the functions $a(u)$, $b(u)$, and $c(u)$), with explicit mass terms that arise through the backreaction of the D8-branes on  the D4 background.  Notice that the action is written in terms of just two physical parameters $M_{\mathrm{KK}}$ and $\zeta$.

\subsection{Expanding the Brane Actions}

Next we turn to the action on the flavor branes, expanding to quadratic order in the scalar $\Phi$ and the bulk graviton components $h^{\tau}_{\mu}$ and $h^{\tau}_{u}$. All contributions come from the DBI action.  The Chern-Simons action, as we argue in appendix \ref{sec:CSappendix}, does not contribute terms of this type.

On the brane, it is natural to use the $Z$ coordinate
which fully covers the D8, $\overline{\text{D8}}$ stacks,
instead of the radial $U$ coordinate.  The scalar $\Phi(x^{\mu}, Z)$ can be thus be thought of as a sum of a field symmetric in $Z$, and a field anti-symmetric in $Z$:
\beq
\Phi = \Phi^{(S)} + \Phi^{(A)} \, .
\eeq

Using the DBI action in equation \eqref{eqn:SDBI} and the pullback of the bulk metric in equation \eqref{eqn:pullback} -- where now the bulk metric includes graviton fluctuations -- we have
\beq
S_{\mathrm{DBI}} \supset -\frac{2\tilde{\zeta} \, \delta\tau}{2\kappa_{10}^2}\int d^8\sigma dZ \, e^{-\tilde{\phi}}\sqrt{\frac{-\det\check{g}}{\tilde{g}_{\tau\tau}}} \, \tilde{g}_{\tau\tau}\left[\tilde{g}^{\mu\nu}h^{\tau}_{\mu}\partial_{\nu}\Phi - \frac{dZ}{dU}\tilde{g}^{UU}h^{\tau}_{U}\partial_Z\Phi + \cdots\right] \, ,
\eeq
where $\det\check{g}$ represents the determinant of the metric expressed in terms of the $Z$-coordinate.  From this form, along with the facts that $\frac{dZ}{dU}$ is an anti-symmetric function of $Z$ and $\{h^{\tau}_{\mu}, h^{\tau}_{U}\}$ must be symmetric functions of $Z$, we can see that only the symmetric part of the scalar $\Phi$ contributes to the coupling.  We will therefore ignore the anti-symmetric part in what follows.\footnote{Note that our $\Phi$ is related by a factor of $Z$ to the brane scalar $y$ in WSS, so the symmetric mode of $\Phi$ in fact corresponds to an antisymmetric, parity-odd mode of $y$.}


We ultimately prefer to work with an integral over $U$ (thinking of all fields as functions of $U$).  The fact that we are working with functions symmetric in $Z$ thus simply determines the boundary condition at $U = U_{\mathrm{KK}}$ (or $u=1$) to be
\beq
0 = \partial_Z\Phi^{(S)}\Big|_{Z = 0} = \frac{dU}{dZ}\partial_{U}\Phi^{(S)}\Bigg|_{U = U_{\mathrm{KK}}} \, .
\eeq
With this, we arrive at the expansion to quadratic order given by

\beq
S_{\mathrm{DBI}} \supset \frac{2\tilde{\zeta}\delta\tau}{2\kappa_{10}^2}\int d^{8}\sigma \, dU \, e^{-\tilde{\phi}}\sqrt{\frac{-\det\tilde{g}}{\tilde{g}_{\tau\tau}}}\tilde{g}_{\tau\tau}
\eeq

$$
\times \left[ - \frac{1}{2}\tilde{g}^{\mu\nu}\partial_{\mu}\Phi^{(S)}\partial_{\nu}\Phi^{(S)} -  \frac{1}{2}\tilde{g}^{UU}\partial_{U}\Phi^{(S)}\partial_{U}\Phi^{(S)}  - \tilde{g}^{\mu\nu}h^{\tau}_{\mu}\partial_{\nu}\Phi^{(S)} - \tilde{g}^{UU}h^{\tau}_{U}\partial_{U}\Phi^{(S)}
\right] \, .
$$

Integrating this expression over the 4-sphere, converting to the dimensionless radial coordinate $u$, and applying the same definitions utilized for the bulk action gives us
\beq
\label{eqn:SDBIclean}
S_{\mathrm{DBI}} \supset K\int d^4x \, du  \, \zeta M_{\mathrm{KK}}^2 c(u)
\eeq

$$
\times \left[- \frac{1}{2}\eta^{\mu\nu}\partial_{\mu}\Phi^{(S)}\partial_{\nu}\Phi^{(S)} -  \frac{1}{2}M_{\mathrm{KK}}^2b(u)\partial_{u}\Phi^{(S)}\partial_{u}\Phi^{(S)}  - \eta^{\mu\nu}h^{\tau}_{\mu}\partial_{\nu}\Phi^{(S)} - M_{\mathrm{KK}}^2b(u)h^{\tau}_{u}\partial_{u}\Phi^{(S)}
\right] \, .
$$

Notice that this action has the same overall constant $K$ as the bulk action, involves the same functions $\{b(u), c(u)\}$ which appeared in our bulk action expansion, and no others, and again depends on the two essential parameters $M_{\mathrm{KK}}$ and $\zeta$.

\subsection{Gauge Symmetry and Absorption of a Scalar Field}\label{sec:gaugesymmetry}

The full quadratic action, the sum of equations \eqref{eqn:Sbulkclean} and \eqref{eqn:SDBIclean}, can be written as

\beq
S = K\int d^4x \, du 
\eeq

$$
\Bigg\{-\frac{a(u)}{4}\eta^{\mu\rho}\eta^{\nu\sigma}\Big(\partial_{\rho}h^{\tau}_{\sigma} - \partial_{\sigma}h^{\tau}_{\rho}\Big)\Big(\partial_{\mu}h^{\tau}_{\nu} - \partial_{\nu}h^{\tau}_{\mu}\Big)
 -\frac{M_{\mathrm{KK}}^2a(u)b(u)\eta^{\mu\nu}}{2}\Big(\partial_{\mu}h^{\tau}_{u} - \partial_{u}h^{\tau}_{\mu}\Big)\Big(\partial_{\nu}h^{\tau}_{u} - \partial_{u}h^{\tau}_{\nu}\Big)
$$

$$
-\frac{\zeta M_{\mathrm{KK}}^2}{2}c(u)\eta^{\mu\nu}\Big(h^{\tau}_{\mu} + \partial_{\mu}\Phi^{(S)}\Big)\Big(h^{\tau}_{\nu} + \partial_{\nu}\Phi^{(S)}\Big)
-\frac{\zeta M_{\mathrm{KK}}^4}{2}c(u)b(u)\Big(h^{\tau}_{u} + \partial_{u}\Phi^{(S)}\Big)^2  \Bigg\} \, .
$$

Suppose for a moment we ignore the D-branes (or equivalently set $\zeta = 0$).  In that case, the bulk fields decouple from the brane modes, and the  object $\{h^{\tau}_{\mu}, h^{\tau}_{u}\}$ is a 5-dimensional massless vector field in a curved background, with a gauge symmetry that was originally a diffeomorphism of the metric, associated with transformations of the $\tau$-direction.  Once we include the D8-branes, this transformation also involves a shift of the scalar field $\Phi$, which is associated with the location of the flavor branes on the $\tau$ circle.  Specifically, the new gauge transformation (really a Stueckelberg-like field redefinition) is
\beq
h^{\tau}_{\mu} \rightarrow h^{\tau}_{\mu} - \partial_{\mu}\xi \, , \hspace{.5in} h^{\tau}_{u} \rightarrow h^{\tau}_{u} - \partial_{u}\xi \, , \hspace{.5in} \Phi^{(S)} \rightarrow \Phi^{(S)} + \xi \, ,
\eeq
where $\xi = \xi(x^{\mu}, u)$.  Note that this is only a valid symmetry because we have included both the backreacted background metric and the mixing term; if we attempt to include either one without the other, we end up with an action that appears to violate this symmetry, and results in a unphysical Lagrangian.  This emphasizes the fact that the mixing term ought, in fact, to be thought of as part of the backreaction.

We can take advantage of this  symmetry to absorb away the scalar $h^{\tau}_{u}$, simplifying our analysis.  (Note that unlike the vector field on the brane analyzed by \cite{Sakai2005}, this vector field includes no massless zero mode to complicate the process.)  We begin by defining the field $\kappa$ such that
\beq
\partial_{u}\kappa = h^{\tau}_{u} \, , 
\eeq
and then we define a vector field $\kappa_{\mu}$ such that
\beq
\kappa_{\mu} = h^{\tau}_{\mu} - \partial_{\mu}\kappa \, , \hspace{.75in} 
F_{\mu\nu} = \partial_{\mu}\kappa_{\nu} - \partial_{\nu}\kappa_{\mu} \, .
\eeq
Finally, we also define the scalar field
\beq
\omega = \sqrt{\zeta}M_{\mathrm{KK}}\Big(\Phi^{(S)} + \kappa\Big) \, .
\eeq

These definitions leave us with 
\beq\label{eq:fullaction}
S = K\int d^4x \, du ~
 \Bigg\{
-\frac{a(u)}{4}\eta^{\mu\rho}\eta^{\nu\sigma}F_{\mu\nu}F_{\rho\sigma}
 -\frac{M_{\mathrm{KK}}^2a(u)b(u)\eta^{\mu\nu}}{2}\partial_{u}\kappa_{\mu}\partial_{u}\kappa_{\nu} - \frac{M_{\mathrm{KK}}^2\zeta c(u)}{2}\eta^{\mu\nu}\kappa_{\mu}\kappa_{\nu}
\eeq

$$
 - \frac{c(u)}{2}\eta^{\mu\nu}\partial_{\mu}\omega\partial_{\nu}\omega - \frac{M_{\mathrm{KK}}^2b(u)c(u)}{2}\partial_{u}\omega\partial_{u}\omega
 - M_{\mathrm{KK}}\sqrt{\zeta}c(u)\eta^{\mu\nu}\kappa_{\mu}\partial_{\nu}\omega \Bigg\} \, ,
$$
an action involving only a 4-component massive vector field $\kappa^{\mu}$ and a scalar $\omega$.

\section{Mode Expansion in the Radial Coordinate}\label{sec:KK}

To determine the mass spectra of the hadrons dual to $\kappa_\mu$ and $\omega$, we first decompose these fields into eigenmodes along the radial direction, $U$. 

Our quadratic action \eqref{eq:fullaction} is a sum of three terms:
\beq
S = S_{S} + S_{V} + S_{M}
\eeq
where $S_{V}$ is the part of the action just involving the vector, $S_{S}$ is the part  just involving the scalar, and $S_{M}$ is the part of the action with the term that mixes the vector with the scalar.

We begin by constructing mode expansions for the vector and scalar fields separately, and expressing the mixing term in terms of these expansions.  Then, we show that the mixing term has no effect on the mass spectrum of the vector glueball, since we can eliminate the mixing by shifting the vector fields by a term akin to a gauge transformation.  This will result in an action for the scalar fields with kinetic and mass terms which cannot be simultaneously diagonalized.  Nonetheless, it is possible to diagonalize the equations of motion, and thus find the spectrum of mass poles for the scalar fields.

\subsection{Vector Mode Expansion}

We begin with the terms quadratic in the vector field:
\beq
S_{V} = K\int d^4x \, du \,  \, \Bigg\{-\frac{a(u)}{4}\eta^{\mu\rho}\eta^{\nu\sigma}F_{\mu\rho}F_{\rho\sigma}
 -\frac{M_{\mathrm{KK}}^2a(u)b(u)\eta^{\mu\nu}}{2}\partial_{u}\kappa_{\mu}\partial_{u}\kappa_{\nu} - \frac{M_{\mathrm{KK}}^2\zeta c(u)}{2}\eta^{\mu\nu}\kappa_{\mu}\kappa_{\nu}\Bigg\} \, ,
\eeq

and expand the vector field in eigenmodes $\psi_n(u)$:
\beq
\kappa_{\mu}(x^{\nu}, u) = \sum_{n} \frac{B^{n}_{\mu}(x^{\nu})\psi_n(u)}{\sqrt{a(u)}} \, .
\eeq
We include the factor of $\frac{1}{\sqrt{a(u)}}$ in this expression so that the ``wavefunctions'' $\psi_n(u)$ will be canonically orthonormal (with a trivial metric).  Defining $f_{\mu\nu}^{n}$ as the field strength of $B^{n}_{\mu}$ and integrating by parts on $u$, we can then write the action in terms of a tower of 4d vector glueballs $\{ B^{n}_{\mu}\}$ as
\beq
S_V = K\int d^4x \, du \, \sum_{n,m} \psi_n\Bigg\{-\frac{\psi_m}{2} \eta^{\mu\rho}\eta^{\nu\sigma}f^{n}_{\mu\nu}f^{m}_{\rho\sigma}
\eeq
$$
+ \frac{M_{\mathrm{KK}}^2}{2}\left(\frac{1}{\sqrt{a(u)}}\partial_{u}\left[a(u)b(u)\partial_{u}\left(\frac{\psi_m}{\sqrt{a(u)}}\right)\right] - \frac{\zeta c(u)}{a(u)}\psi_m\right)\eta^{\mu\nu}B^{n}_{\mu}B^{m}_{\nu}\Bigg\} \, .
$$

The $\psi_n(u)$ can be chosen to be eigenstates of the operator $\hat{H}_{\psi}$ such that
\beq
\hat{H}_{\psi}\psi_n = -\frac{1}{\sqrt{a(u)}}\partial_{u}\left[a(u)b(u)\partial_{u}\left(\frac{\psi_n}{\sqrt{a(u)}}\right)\right] + \frac{\zeta c(u)}{a(u)}\psi_n = \lambda_{n}\psi_n \, ,
\eeq
with which our vector action becomes
\beq
S_V = K\int d^4x \, \sum_{n} \Bigg\{-\frac{1}{2}\eta^{\mu\rho}\eta^{\nu\sigma}f^{n}_{\mu\nu}f^{n}_{\rho\sigma} - \frac{M_{\mathrm{KK}}^2\lambda_{n}}{2}\eta^{\mu\nu}B^{n}_{\mu}B^{n}_{\nu}\Bigg\} \, .
\eeq

In what follows, we will also use notation where $\psi_n(u)$ is represented by the ket $|n\rangle$. In this language,
\beq
\hat{H}_{\psi}|n\rangle = \lambda_{n}|n\rangle \, , \hspace{.75in}
\langle n | m \rangle = \delta_{nm} \, , \hspace{.75in} \sum_{n} |n\rangle\langle n | = \hat{1} \, .
\eeq

\subsection{Scalar Mode Expansion}

Now consider the  scalar-only part of the  action
\beq
S_S = K\int d^4x \, du \, c(u) \, \Bigg\{ - \frac{1}{2}\eta^{\mu\nu}\partial_{\mu}\omega\partial_{\nu}\omega - \frac{M_{\mathrm{KK}}^2b(u)}{2}\partial_{u}\omega\partial_{u}\omega\Bigg\} \, .
\eeq
Again, we assume we have an expansion for the scalar field of the form
\beq
\omega(x^{\mu}, u) = \sum_{i} \frac{\varpi^{i}(x^{\mu})\varphi_i(u)}{\sqrt{c(u)}} \, ,
\eeq
where the factor $\frac{1}{\sqrt{c(u)}}$ is included so that the $\varphi_i(u)$ wavefunctions are canonically orthonormal.  

This allows us to rewrite the scalar action in terms of the scalar tower $\{\varpi^{i}\}$ as
\beq
S_{S} = K\int d^4x \, du \, \sum_{i,j} \, \varphi_i \, \Bigg\{-\varphi_j\eta^{\mu\nu}\partial_{\mu}\varpi^{i}\partial_{\nu}\varpi^{j} + \frac{M_{\mathrm{KK}}^2}{2}\frac{1}{\sqrt{c(u)}}\partial_{u}\left[b(u)c(u)\partial_u\left(\frac{\varphi_j}{\sqrt{c(u)}}\right)\right]\varpi^{i}\varpi^{j}\Bigg\} \, .
\eeq

Here we define an operator $\hat{H}_{\varphi}$ whose eigenstates are $\varphi_i(u)$ with
\beq
\hat{H}_{\varphi}\varphi_i = -\frac{1}{\sqrt{c(u)}}\partial_{u}\left[b(u)c(u)\partial_u\left(\frac{\varphi_i}{\sqrt{c(u)}}\right)\right] = \chi_i\varphi_i(u) \, .
\eeq
Our scalar action becomes
\beq
S_{S} = K\int d^4x \sum_{i} \Bigg\{-\frac{1}{2}\eta^{\mu\nu}\partial_{\mu}\varpi^{i}\partial_{\nu}\varpi^{i} - \frac{M_{\mathrm{KK}}^2\chi_i}{2}\Big(\varpi^{i}\Big)^2\Bigg\} \, .
\eeq
Representing the wavefunction $\varphi_i(u)$ as $|i\rangle$, we define a similar notation for the scalars as we did for the vectors:
\beq
\hat{H}_{\varphi}|i\rangle = \chi_i|i\rangle \, , \hspace{.75in} \langle i | j \rangle = \delta_{ij} \, , \hspace{.75in} \sum_{i} |i\rangle\langle i | = \hat{1} \, .
\eeq 
Here $\chi_i$ is the eigenvalue of the $i$th state.
Note that the sets $\{\psi_n(u)\}$ and $\{\varphi_i(u)\}$ will be distinct orthonormal bases for the same Hilbert space.

\subsection{Mass Eigenstates for the Bulk Vector and Brane Scalar Modes}

We now turn to the mixing term, 
\beq
S_M = K\int d^4x \, du \Bigg\{-M_{\mathrm{KK}}\sqrt{\zeta} c(u)\eta^{\mu\nu}\kappa_{\mu}\partial_{\nu}\omega\Bigg\} \, .
\eeq
If we expand this in terms of our mode towers for the vector and scalar, and define the ``mixing operator''
\beq
\hat{\mathcal{K}} \, \psi(u) = \sqrt{\frac{c(u)}{a(u)}} \, \psi(u) \, ,
\eeq
we can write this as
\beq
S_M = K\int d^4x \, \sum_{i,n}\Bigg\{-M_{\mathrm{KK}}\sqrt{\zeta}\langle i | \hat{\mathcal{K}} | n \rangle \eta^{\mu\nu}B^{n}_{\mu}\partial_{\nu}\varpi^{i}\Bigg\} \, .
\eeq

When we combine this with the previously analyzed vector action, we have
\beq
S_V + S_M = K\int d^4x \, \sum_{i, n}\Bigg\{-\frac{1}{4}\eta^{\mu\rho}\eta^{\nu\sigma}f^{n}_{\mu\nu}f^{n}_{\rho\sigma} 
\eeq
$$
- \frac{M_{\mathrm{KK}}^2\lambda_{n}}{2}\eta^{\mu\nu}\left[B^{n}_{\mu}B^{n}_{\nu} + \frac{2\sqrt{\zeta}}{\lambda_{n}M_{\mathrm{KK}}}\sum_{i} \langle i | \hat{\mathcal{K}} | n \rangle B^{n}_{\mu}\partial_{\nu}\varpi^{i}\right]\Bigg\} \, .
$$
We can now define a shift of the vector tower which ``completes the square'' to eliminate the mixing between scalars and vectors.  This is
\beq
\check{B}^{n}_{\mu} = B_{\mu}^{n} + \frac{\sqrt{\zeta}}{\lambda_{n}M_{\mathrm{KK}}}\sum_{i} \langle i | \hat{\mathcal{K}} | n \rangle\partial_{\mu}\varpi^{i} \, ,
\eeq
noting that the field strengths of the vector fields satisfy $f^{n}_{\mu\nu} = \check{f}^{n}_{\mu\nu}$.  As a result we can write $S_V + S_M = \check{S}_V + \check{S}_M$ with
\beq
\check{S}_V = K\int d^4x \, \sum_{n} \Bigg\{-\frac{1}{4}\eta^{\mu\rho}\eta^{\nu\sigma}\check{f}^{n}_{\mu\nu}\check{f}^{n}_{\rho\sigma} - \frac{M_{\mathrm{KK}}^2\lambda_{n}}{2}\eta^{\mu\nu}\check{B}^{n}_{\mu}\check{B}^{n}_{\nu}\Bigg\} \, ,
\eeq
and
\beq
\check{S}_{M} = K\int d^4x \, \sum_{i,j,n} \Bigg\{\frac{\zeta}{2\lambda_{n}}\langle i | \hat{\mathcal{K}} | n \rangle\langle n | \hat{\mathcal{K}} | j \rangle \eta^{\mu\nu}\partial_{\mu}\varpi^{i}\partial_{\nu}\varpi^{j}\Bigg\}
 = K\int d^4x \, \sum_{i,j} \Bigg\{\frac{\zeta}{2}\langle i | \hat{\mathcal{K}}\hat{H}^{-1}_{\psi}\hat{\mathcal{K}}| j \rangle \eta^{\mu\nu}\partial_{\mu}\varpi^{i}\partial_{\nu}\varpi^{j}\Bigg\} \, .
\eeq

At this stage we have decoupled the vector field. Note that the mass spectrum for the vector will simply be determined by the eigenvalues of the operator $\hat{H}_{\psi}$: although the backreaction of the branes onto the background metric has an effect on the masses, the mixing term itself does not.

The mixing term does, however, have a significant impact on the scalar spectrum. Assembling the remaining terms into a new action for the scalar fields, we can write
\beq
\check{S}_S = S_S + \check{S}_{M} = K\int d^4x \, \sum_{i,j} \, \Bigg\{-\frac{1}{2}\left\langle i \left|\Big(\hat{1} - \zeta\hat{\mathcal{K}}\hat{H}_{\psi}^{-1}\hat{\mathcal{K}}\Big)\right| j \right\rangle\eta^{\mu\nu}\partial_{\mu}\varpi^{i}\partial_{\nu}\varpi^{j} - \frac{M_{\mathrm{KK}}^2\langle i | \hat{H}_{\varphi} | j \rangle}{2}\varpi^{i}\varpi^{j}\Bigg\} \, .
\eeq

The matrices multiplying the kinetic and mass terms here are not simultaneously diagonalizable, so we cannot rewrite this action as a sum of separate scalar actions.  But if instead we work with the equations of motion, we can say
\beq
\partial^2\varpi^{i} - M_{\mathrm{KK}}^2\sum_{j}\left\langle i \left|\hat{H}_{\check{\varphi}}\right| j \right\rangle\varpi^{j} = 0 \, ,
\eeq
with
\beq\label{eqn:Hphi}
\hat{H}_{\check{\varphi}}|j\rangle = \Big(\hat{1} - \zeta\hat{\mathcal{K}}\hat{H}_{\psi}^{-1}\hat{\mathcal{K}}\Big)^{-1}\hat{H}_{\varphi}\Big| j \Big\rangle \, .
\eeq
The mass poles of the scalar fields are then given by eigenvalues $\check{\chi}_i$ of the matrix $\langle i | \hat{H}_{\check{\varphi}} | j \rangle$.  

Note that the operator $\hat{H}_{\check{\varphi}}$ is not Hermitian, which is potentially an issue, since it ought to be associated with an observable in this system.  However, we should remember that our entire construction, starting from equation \eqref{eqn:BREOM}, is only valid to linear order in $\zeta$, because we have added the D8-branes in after taking the supergravity limit of the background.  It is possible that a complete treatment of the brane system would remedy this problem.  In any case, to linear order in zeta we can use standard perturbation theory to find the eigenvalues, while leaving the eigenstates unaffected.

\section{Linear Order Results}\label{sec:results}

From this point on we will be working to linear order in $\zeta$.  Essentially, we need to expand the operators $\hat{H}_{\check{\varphi}}$ and $\hat{H}_{\psi}$ to linear order in $\zeta$, and use first order perturbation theory to compute the leading correction to the eigenvalue spectra.  Note that in addition to the explicit $\zeta$ dependence in both operators, we also have the functions $a(u)$, $b(u)$, and $c(u)$, which are determined by the perturbed background, and therefore need to be expanded in $\zeta$.  In the case of the scalar field mass spectrum, we will keep separate the perturbation to  arising from the explicit mixing term from that arising from perturbations to the background metric in order to analyze their impacts individually.

\subsection{Expanding the Hamiltonians}

We will begin by assuming the functions $a(u)$, $b(u)$, and $c(u)$ take the form

\beq
a(u) = a^{(0)}(u)\Big(1 + \zeta a^{(1)}(u)\Big) \, , \hspace{.5in} b(u) = b^{(0)}(u)\Big(1 + \zeta b^{(1)}(u)\Big) \, , \hspace{.5in} c(u) = c^{(0)}(u)\Big(1 + \zeta c^{(1)}(u)\Big) \, .
\eeq

In this case, when we expand our vector operator out to linear order, we obtain
\beq
\hat{H}_{\psi} = \hat{H}_{\psi}^{(0)} + \zeta \, \delta \hat{H}_{\psi} \, , 
\eeq
where
\beq
\hat{H}^{(0)}_{\psi}\psi = -\frac{1}{\sqrt{a^{(0)}}}\partial_{u}\left[a^{(0)}b^{(0)}\partial_{u}\left(\frac{\psi}{\sqrt{a^{(0)}}}\right)\right] \, , 
\eeq
and
\beq
\delta \hat{H}_{\psi}\psi = - \frac{\partial}{\partial u}\left[b^{(0)}b^{(1)}\frac{\partial\psi}{\partial u}\right] + \Bigg[\frac{1}{2\sqrt{a^{(0)}}}\frac{\partial}{\partial u}\left(\frac{a^{(0)}{}'b^{(0)}b^{(1)}}{\sqrt{a^{(0)}}}\right) + \frac{1}{2a^{(0)}}\frac{\partial}{\partial u}\Big(a^{(0)}a^{(1)}{}'b^{(0)}\Big) + \frac{c^{(0)}}{a^{(0)}}\Bigg]\psi \, .
\eeq

We can write the eigenstates of the unperturbed Hamiltonian as $|n\rangle_0$, so that
\beq
\hat{H}^{(0)}_{\psi}|n\rangle_0 = \lambda_n^{(0)}|n\rangle_0 \, .
\eeq
Then, using first order perturbation theory, we find the linear corrections to the eigenvalues as
\beq
\lambda_{n} = \lambda_{n}^{(0)} + \zeta\delta\lambda_n = \lambda_{n}^{(0)} + \zeta\langle n | \delta \hat{H}_{\psi} | n \rangle_{0} \, .
\eeq

Moving on to our scalar Hamiltonian, we expand to linear order and obtain
\beq
\hat{H}_{\check{\varphi}} = \hat{H}_{\varphi}^{(0)} + \zeta \Big(\delta \hat{H}_{\check{\varphi}, 1} + \delta\hat{H}_{\check{\varphi}, 2}\Big) \, , 
\eeq

where
\beq
\hat{H}^{(0)}_{\varphi}\varphi = -\frac{1}{\sqrt{c^{(0)}}}\partial_{u}\left[c^{(0)}b^{(0)}\partial_{u}\left(\frac{\varphi}{\sqrt{c^{(0)}}}\right)\right] \, , 
\eeq
and
\beq
\delta \hat{H}_{\check{\varphi}, 1} \, \varphi = - \frac{\partial}{\partial u}\left[b^{(0)}b^{(1)}\frac{\partial\varphi}{\partial u}\right] + \Bigg[\frac{1}{2\sqrt{c^{(0)}}}\frac{\partial}{\partial u}\left(\frac{c^{(0)}{}'b^{(0)}b^{(1)}}{\sqrt{c^{(0)}}}\right) + \frac{1}{2c^{(0)}}\frac{\partial}{\partial u}\Big(c^{(0)}c^{(1)}{}'b^{(0)}\Big)\Bigg]\varphi \, ,
\eeq
and
\beq
\label{eqn:dHvarphi2}
\delta \hat{H}_{\check{\varphi}, 2} \, \varphi =  \hat{\mathcal{K}}^{(0)}\Big(\hat{H}_{\psi}^{(0)}\Big)^{-1}\hat{\mathcal{K}}^{(0)}\hat{H}_{\varphi}^{(0)}\varphi \, .
\eeq

The eigenstates of the unperturbed Hamiltonian in this case are $|i\rangle_0$ with
\beq
\hat{H}^{(0)}_{\varphi}|i\rangle_0 = \chi_i^{(0)}|i\rangle_0 \, ,
\eeq

and first order perturbation theory gives us
\beq
\chi_i = \chi_{i}^{(0)} + \zeta\Big(\delta\chi_{i, 1} + \delta\chi_{i,2}\Big) = \chi_{i}^{(0)} + \zeta\Big(\langle i | \delta \hat{H}_{\check{\varphi}, 1} | i \rangle_{0} + \langle i | \delta\hat{H}_{\check{\varphi}, 2}| i \rangle_{0}\Big) \, .
\eeq

In order to determine the functions $a^{(1)}(u)$, $b^{(1)}(u)$, and $c^{(1)}(u)$, we used the work of Bigazzi et al. \cite{Bigazzi2015},
which established the corrections to the metric up to linear order in $\zeta$, summarized in appendix \ref{sec:LinearBackreactionApp}.  The authors showed that these linear-order perturbation functions satisfy second order differential equations with closed form solutions in terms of hypergeometric functions.  The constants of integration appearing there were fixed by requiring regularity at $U = U_{\mathrm{KK}}$ and removing the most divergent terms in $U \rightarrow \infty$, along with other restrictions. In this work, we choose to impose the most stringent requirements discussed in \cite{Bigazzi2015} rather than just those needed for satisfying physical constraints.  This is done primarily to provide concrete solutions which we can use in the next section.  

\begin{figure}
\begin{center}
\includegraphics{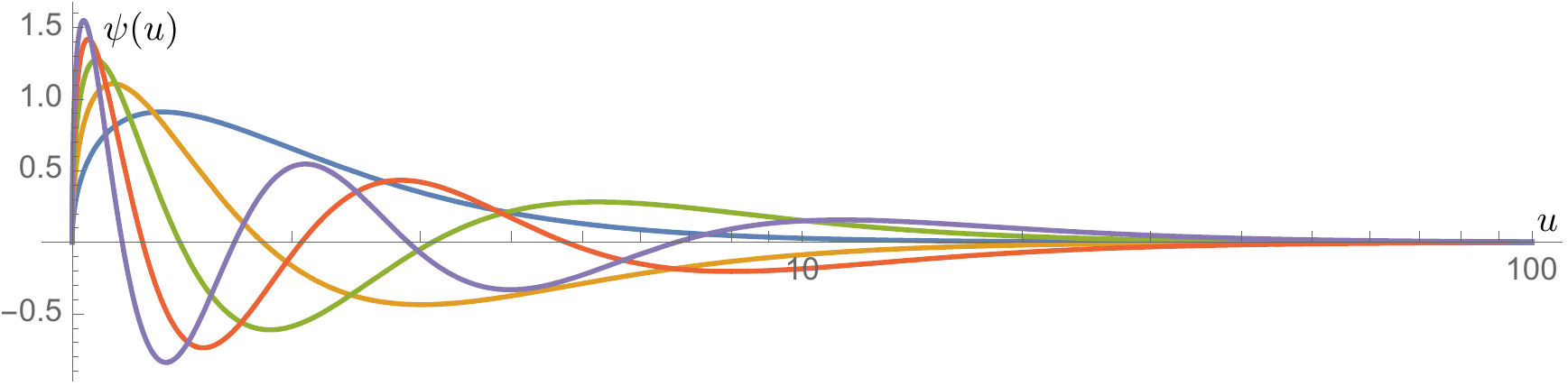}
\caption{\label{vectorpic} The lowest five unperturbed vector functions $\psi_n(u)$.}
\end{center}
\end{figure}

\begin{figure}
\begin{center}
\includegraphics{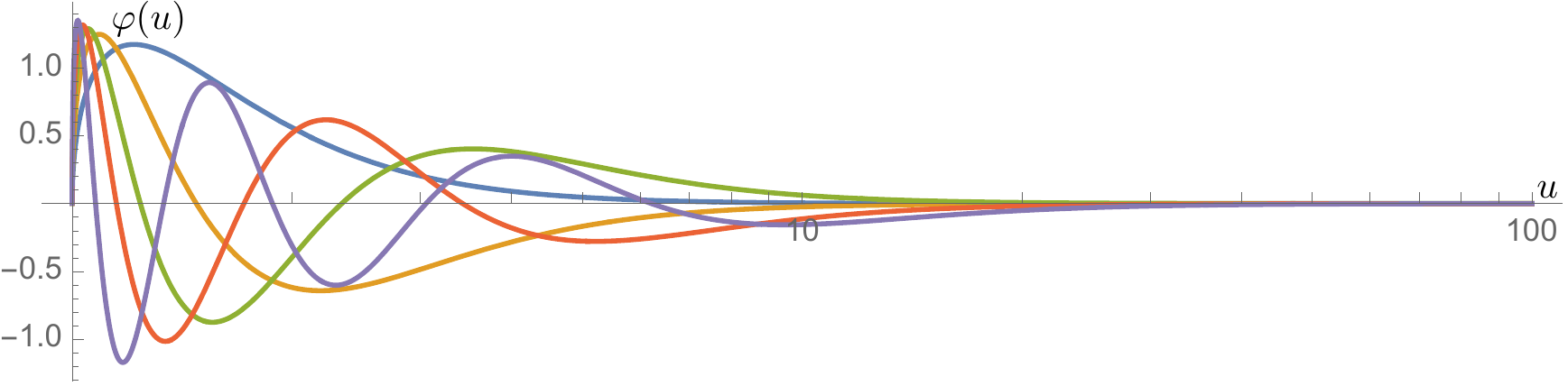}
\caption{\label{scalarpic} The lowest five unperturbed scalar functions $\varphi_j(u)$.}
\end{center}
\end{figure}

It is also worth noting that these solutions can only be trusted in the regime $u \ll \frac{1}{\zeta}$, as noted in \cite{Burrington2008}.  In our context, this means using only those unperturbed states $|n\rangle_0$ and $|j\rangle_0$ which die off by the time $u \sim \frac{1}{\zeta}$ (forcing us to restrict our analysis to the lowest members of each mode tower, or choose to assume an extremely small value of $\zeta$, or both).  Figures \ref{vectorpic} and \ref{scalarpic} show the lowest five unperturbed wavefunctions associated with both the vectors and scalars. These wavefunctions are very small for $u> 100$.  As a result, in order for our results to be trusted rigorously, we must assume $\zeta \ll 0.01$, a substantially smaller value than the choice $\zeta = 0.33$ obtained by fixing the parameters to $m_{\rho}$ and $f_{\pi}$.  

\subsection{Discussion of Numerical Results}

In order to generate our numerical results, we implemented the ``shooting method'' using \texttt{Mathematica}, utilizing built-in differential equation solvers to find the unperturbed wavefunctions.  Then, we computed the overlap integrals necessary to find the perturbations of the eigenvalues using built-in numerical integration tools.  The details of this process, together with the numerical parameters used, are given in appendix \ref{sec:NUMappendix}.

The numerical results for the eigenvalues corresponding to mixed glueball and meson states are reported in Table \ref{tab:lambdas}. The eigenvalues $\lambda_n$ correspond to the glue tower, while the $\chi_j$ correspond to mesons -- though the corrected mass eigenstates are, of course, admixtures of glueball and meson. The glueball eigenvalues are related to the masses as
\begin{align}
    m_n^2=(\lambda_n+\zeta\delta\lambda_n) M_{KK}^2
\end{align}
and similarly for the meson masses with $\chi_j$.

Due to the structure of the glueball-meson coupling from DBI action, the corrections to the glueball tower come entirely from the background metric, and are negative (pushing the masses downward) for all states. This aligns with results for vector and axial-vector meson masses discussed in \cite{Bigazzi2015}. 

The meson tower (now with glueball admixtures) is more interesting: while the effect of the modified background geometry is still to push the eigenvalues downward, the contribution from the mixing term is positive. For the four lightest mesons, the mixing term is strong enough to make the overall correction positive, but is overtaken by the background contribution for the fifth and higher modes. Indeed, the unperturbed wavefunctions $\psi$ and $\varphi$ have the greatest overlap for lighter states, and have little overlap for heavier states. A similar effect was observed in \cite{Rinaldi2020PureGS} for in a bottom-up holographic model. The net effect of the corrections (including both background and mixing term) is to decrease the slope of roughly linear dependence of $m^2$ on excitation number. 

Both of the towers we are analyzing are considered spurious, based on parity and charge-conjugation quantum numbers. We believe that the vector mode of the graviton corresponds to a $J^{PC}=1^{+-}$ glueball, identified in Brower et al. \cite{BROWER2000} as a spurious state, based on comparison to lattice data in the quenched approximation \cite{Athenodorou:2020ani}. We should remark that \cite{BROWER2000}  identified this mode as a $1^{-+}$ glueball. We believe that there is a some ambiguity in the parity assignments, however, as noted also in \cite{Imoto:2010ef}. \cite{BROWER2000} investigates only bulk modes in the wrapped D4-brane background, and notes that there are two separate parity operations, $P:~x^i\rightarrow -x^i$ and $P_\tau:~\tau\rightarrow -\tau$, and bases 4d parity assignments on the action of $P$, with $P_\tau$ acting as a spurious additional symmetry. In \cite{Sakai2005}, with the introduction of flavor branes, this ambiguity is resolved because the larger symmetry group is broken to a subgroup that takes $(x^i,\tau)\rightarrow (-x^i,-\tau)$, which is thus unambiguously equivalent to 4d parity for the purpose of identifying the 4d quantum numbers of a given mode. Based on this definition, the vector glueball we consider, which comes from the graviton mode $h^\tau_\mu$, has $P_\tau=-1$, meaning that it violates natural parity and should be identified as a $P=+$ state. Similarly, as noted in \cite{Sakai2005} the action of charge conjugation is to flip the orientation of strings and act with $P_\tau$. Thus the glueball state we consider should be a $1^{+-}$ mode.

The brane scalar we study corresponds to a isospin 0, $0^{--}$. In the original WSS model, this state is the second-lightest excitation of the transverse fluctuation mode, as noted by \cite{Sakai2005}, identified there to be parity odd.\footnote{In this work, this is the lightest mode of the {\em symmetric} part of the scalar $\Phi$, which is, however, related to the scalar field $y$ of \cite{Sakai2005} by a factor of $Z$.}

By including the leading $N_f/N_c$ corrections to the WSS model, we are making the model's predictions more precise. It is an important question whether this improves or worsens the alignment of WSS with experimental and lattice data.
The light $0^{--}$ states are famously artifacts of holographic QCD models, as no such states appear in the light spectrum. It is thus encouraging, that here their masses are pushed higher. On the other hand, the masses of the $1^{+-}$ glueballs (also believe to be spurious) decrease, which is less promising.  

\begin{table}\label{tab:lambdas}
\begin{center}
\begin{tabular}{|c|c|c|}
\hline
$n$ & $\lambda_n$ & $\delta\lambda_n$ \\
\hline
\hline
1 & 3.554 & -1.280 \\
2 & 8.053 & -4.106 \\
3 & 14.016 & -8.652 \\
4 & 21.459 & -15.069 \\
5 & 30.387 & -23.477 \\
\hline
\end{tabular}
\hspace{.5in}
\begin{tabular}{|c|c|c|c|c|}
\hline
$j$ & $\chi_j$ & $\delta\chi_{j,1}$ & $\delta\chi_{j,2}$ & $\delta\chi_{j}$ \\
\hline
\hline
1 & 5.310 & -0.932 & 3.855 & 2.923 \\
2 & 10.5354 & -4.207 & 7.106 & 2.899 \\
3 & 17.2392 & -9.098 & 11.280 & 2.182 \\
4 & 25.426 & -15.780 & 16.371 & 0.591 \\
5 & 46.255 & -34.987 & 29.299 & -5.688 \\
\hline
\end{tabular}
\end{center}
\caption{\label{vectortable} On the left, a table of numerical values for the eigenvalues $\lambda_n$ in the unperturbed background as well as the correction terms $\delta\lambda_n$. These are related to the mass of the vector glueball as $m^2=(\lambda+\delta\lambda)M_{KK}^2$.  On the right, a table of numerical values for the unperturbed eigenvalues $\chi_j$ corresponding to the scalar meson, as well as the corrections terms: $\delta\chi_{j, 1}$ due to the mixing, and $\delta\chi_{j, 2}$ due to the perturbed background, and $\delta\chi_j =\delta\chi_{j, 1}+\delta\chi_{j, 2} $.}
\end{table}

\section{Conclusion}\label{sec:conclusion}

We have showed that the WSS model incorporates mixing of glueballs and mesons via the DBI action. As a proof of concept, we demonstrated the effect of this mixing for a sector of the spectrum which decouples from the rest in a smeared-brane approximation of the background: a vector mode of the graviton and the branes' scalar field. 

We found that in order to have a physically sensible effective Lagrangian for these states (and to cancel the bulk-field tadpoles in DBI), one must also include the leading backreaction of the flavor branes on the background. This crucial point is perhaps not surprising in retrospect: while the quadratic term mixing the vector and scalar modes appears even in a naive expansion of the DBI in the un-backreacted geometry, there are additional terms appearing in the effective Lagrangian due to backreaction, which are of the same order in $N_f/N_c$. 
Conversely, the brane scalar -- corresponding to transverse fluctuations of the flavor branes -- plays a crucial role in realizing the ``gauge symmetry'' of the vector graviton mode.

As both the kinetic terms and mass terms in this system cannot be simultaneously diagonalized, we described a general procedure for determining the mass eigenstates -- equivalent to finding the poles of a two-point function --  that lends itself easily to numerical analysis.

We found that while the leading order backreaction of the metric tends to depress meson and glueball masses, the explicit mixing provides a positive contribution which overtakes the metric contribution for low-lying modes. 

There are many interesting directions we hope to explore in the future. First: we worked in an approximation that considered only the trivial mode of the background geometry in the $\tau$ direction, ignoring the (broken) chiral symmetry of the background. The first non-trivial mode -- not suppressed by any parameter compared to the first -- would in fact induce a mixing between the $1^{+-}$ glueball and experimentally relevant $f_0$ mesons, which have long been conjectured to mix with glueball states\cite{ParticleDataGroup:2020ssz}. Including this mode would be very interesting -- but would also, of course, ruin the decoupling of the 4d vector  $h_{\tau\mu}$ from other graviton modes, rendering the analysis significantly more complicated. It would also be interesting see whether trends we observe, like a decrease in the slope of mass-squared versus excitation number for light scalars holds more generally for other mesons. (\cite{Bigazzi2015} observed a similar trend in the predicted spectrum of excited $\rho$ mesons.)

In addition, the type of DBI-induced mixing we observe is ubiquitous in backreacted brane intersections, such as the ones reviewed in \cite{Nunez:2010sf}, or  in the few (supersymmetric) brane intersections like those in \cite{Burrington:2004id, Kirsch:2005uy}, where closed form solutions for localized, un-smeared flavor branes have been found. It would be interesting to understand whether the trends observed in the WSS model -- like the structure of the glueball-meson mixing  which leaves the glueball mass unaffected, or the tendency of the mixing term to overtake the effect of the gravitational backreaction, is in fact universal, or whether some of these effects are unique to Sakai-Sugimoto and/or the smeared approximation.

Finally, the operator technique we used to find the mass eigenvalues (leading to the non-Hermitian ``Hamiltonian" in equation \eqref{eqn:Hphi} was exact in $\zeta$. While this particular problem fixes $\zeta$ to be small (as the background we consider is by definition first order), one might explore the implications of this non-Hermeticity for more general functions $a(u),~b(u),~c(u)$ and larger values of $\zeta$. 

We leave these explorations to future work.

\subsection*{Acknowledgements}
SKD's work is supported by the National Science Foundation (NSF grant number PHY-2014025). NM thanks the KITP and the KITP Scholars Program for hospitality during the final stages of this work, which was supported in part by the National Science Foundation under grant number NSF PHY-1748958. SKD and NM thank Josef Leutgeb and Anton Rebhan for helpful correspondence related to this work. SKD also thanks Ben Burrington and Andy Royston for useful discussions.

\appendix

\section{\label{sec:LinearBackreactionApp} Summary of the Linear Order Backreaction}

Our numerical results rely on using the work of Bigazzi \& Cotrone (BC) \cite{Bigazzi2015}, who established expressions for the perturbations of the background metric and dilaton to linear order in $\zeta$.  We summarize their results here.

BC write the backreacted metric as

\beq
\tilde{g}_{\mu\nu} = e^{2\lambda}\eta_{\mu\nu} \, , \hspace{.5in} \tilde{g}_{\tau\tau} = e^{2\tilde{\lambda}} \, , 
\eeq
\beq
\tilde{g}_{\alpha\beta} = \ell_s^2e^{2\nu}s_{\alpha\beta} \, , \hspace{.5in} \tilde{g}_{\rho\rho} = \ell_s^2e^{-2\varphi} \, , 
\eeq
with
\beq
f(U) = e^{-3r} = \mathrm{Exp}\left[-\frac{3U_{\mathrm{KK}}^3\rho}{\ell_s^3g_s^2}\right] \, ,
\eeq
and the relationship with the dilaton,
\beq
2\phi = \varphi + 4\lambda + \tilde{\lambda} + 4\nu \, .
\eeq
They then expand the functions $\lambda(r)$, $\tilde{\lambda}(r)$, $\nu(r)$, and $\phi(r)$ in the parameter $\epsilon_F = \frac{9\zeta}{4}$, defining
\beq
\lambda(r) = \lambda_0(r) + \epsilon_F\lambda_1(r) + \cdots \, ,
\eeq
and similarly for the other three functions.  By writing down the backreacted equations of motion and integrating out over $\tau$ (see equation \eqref{eqn:BREOM} along with others), BC identified  second order differential equations with source terms for these functions.  These differential equations can be solved in closed form, with
\beq
\lambda_1 = \frac{3}{8}\tilde{f} + y - \frac{1}{4}\Big(A_2 + B_2r\Big) + \frac{1}{4}\Big(A_1 + B_1r\Big) \, , 
\eeq
\beq
\tilde{\lambda}_1 = -\frac{1}{8}\tilde{f} + y - \frac{1}{4}\Big(A_2 + B_2r\Big) - \frac{3}{4}\Big(A_1 + B_1r\Big) \, , 
\eeq
\beq
\phi_1 = \frac{11}{8}\tilde{f} + y - \frac{5}{4}\Big(A_2 + B_2r\Big) + \frac{1}{4}\Big(A_1 + B_1r\Big) \, , 
\eeq
\beq
\nu_1 = \frac{11}{24}\tilde{f} + q \, ,
\eeq
the functions $\{\tilde{f}, y, q\}$ then expressed as
\beq
\tilde{f} = \frac{4}{9}e^{-3r/2} \ _3F_2\left(\frac{1}{2}, \frac{1}{2}, \frac{13}{6}; \frac{3}{2}, \frac{3}{2}; e^{-3r}\right) \, ,
\eeq
\beq
y = z + C_2 - \left[C_1 + C_2\left(1 + \frac{3r}{2}\right)\right]\coth\left(\frac{3r}{2}\right) \, , 
\eeq
\beq
q = 2M_2 + \frac{1}{12}\Big[(A_1 + B_1 r) - 5(A_2 + B_2 r)\Big] + \frac{5}{3}z - \Big[M_1 + M_2(2 + 3r)\Big]\coth\left(\frac{3r}{2}\right) \, ,
\eeq
and finally the function $z$ written as
\beq
z = -\frac{e^{-9r/2}(e^{-3r} + 1)\left[9e^{3r} \ _3F_2\left(\frac{1}{2}, \frac{1}{2}, \frac{19}{6}; \frac{3}{2}, \frac{3}{2}; e^{-3r}\right) \ + \ _3F_2\left(\frac{3}{2}, \frac{3}{2}, \frac{19}{6}; \frac{5}{2}, \frac{5}{2}; e^{-3r}\right)\right]}{162(1 - e^{-3r})} 
\eeq
$$
 - \  \frac{8e^{-3r/2}(10e^{-3r} + 3) \ _2F_1\left(\frac{1}{6}, \frac{1}{2}; \frac{3}{2}; e^{-3r}\right)}{819(1 - e^{-3r})} \ + \ \frac{e^{-15r/2}(38e^{3r} + 8e^{6r} - 40)}{273(1 - e^{-3r})^{13/6}} \, .
 $$
Note that we have added the ``tilde'' to their function $\tilde{f}$, so as to avoid confusion with the function defined in equation \ref{SSstrframe}.  At this stage there are eight unfixed constants of integration: $\{A_1, A_2, B_1, B_2, C_1, C_2, M_1, M_2\}$, and six physical constraints can be imposed.  A zero energy constraint implies
\beq
5B_1 - B_2 - 18(C_2 + 4M_2) = 0 \, ,
\eeq
requiring regularity as $r \rightarrow \infty$ (as $U \rightarrow U_0$) implies
\beq
B_1 = 6C_2 \, , \hspace{.25in} B_2 = 0 \, , \hspace{.25in} M_2 =\frac{C_2}{6} \, ,
\eeq
and although not strictly required, we will also make use of the condition
\beq
C_2 = 0 \, ,
\eeq
which removes all logarithmic divergences as $r \rightarrow \infty$.  

Finally, BC  performed an analysis of the behavior as $r \rightarrow 0$ ($U \rightarrow \infty$), and established that the remaining freedom in the constants of integration would correspond to givings sources or VEVs to other gauge invariant operators, but they also noted that the most divergent terms could be removed with the choices 
\beq
C_1 + C_2 = k \, , \hspace{.5in} M_1 + 2M_2 = \frac{5}{3}k \, ,
\eeq
with
\beq
k = \frac{\pi^{3/2}\Big[3 + \sqrt{3} \, \pi - 12\ln 2 + 9\ln 3\Big]}{78 \, \Gamma\left(-\frac{2}{3}\right) \, \Gamma\left(\frac{1}{6}\right)} \, ,
\eeq
and the next sub-leading divergences with
\beq
A_1 = \frac{81 \, \sqrt{3} \, \pi^{2}\Big[-9 + \sqrt{3} \, \pi - 12\ln 2 + 9 \ln 3\Big]}{43120 \, (2)^{2/3} \, \Gamma\left(-\frac{14}{3}\right) \, \Gamma\left(-\frac{2}{3}\right)^2} \, , \hspace{.75in} A_2 = -2A_1 \, .
\eeq
For concreteness, we have adopted all of these choices, giving us a fully determined solution for the linear perturbation of the background.

Relating this back to the functions $a^{(1)}$, $b^{(1)}$, and $c^{(1)}$ required for our analysis then gives us

\beq
\label{eqn:a1}
a^{(1)} = -\frac{15}{8}\tilde{f} + 9y + 18q - \frac{81A_1}{4} \, ,
\eeq
\beq
\label{eqn:b1}
b^{(1)} = -\frac{3}{8}\tilde{f} - 9y - 18q + \frac{63A_1}{4} \, ,
\eeq
\beq
\label{eqn:c1}
c^{(1)} = \frac{51}{16}\tilde{f} + \frac{27}{2}y + 18q - \frac{81A_1}{8} \, .
\eeq

\section{
\label{sec:BulkActionExpn} Perturbing the Bulk Action to Quadratic Order}

We begin with the bulk action given in equation \eqref{eqn:Sbulk}, expanding it out to quadratic order in excitations around the background $h^{\tau}_{\mu}$ and $h^{\tau}_{U}$.

\subsection{Assumptions and Simplifications}

We work with a backreacted background, but this backreaction treats the D8-branes as ``smeared out'' in the $\tau$-direction, as discussed in \cite{Bigazzi2015}.  This implies that we have the structure

\beq
ds^2 = \tilde{g}_{\mu\nu}dx^{\mu}dx^{\nu} + \tilde{g}_{\tau\tau}d\tau^2 + \tilde{g}_{UU}dU^2 + \tilde{g}_{\alpha\beta}dx^{\alpha}dx^{\beta}
\eeq

with 
\beq
\tilde{g}_{\mu\nu} = \eta_{\mu\nu} \times \Big(\mbox{a function of U}\Big)
\eeq
\beq
\tilde{g}_{\tau\tau} = \mbox{a function of U}
\eeq
\beq
\tilde{g}_{UU} = \mbox{a function of U}
\eeq
\beq
\tilde{g}_{\alpha\beta} = s_{\alpha\beta} \times \Big(\mbox{a function of U}\Big)
\eeq
(where $s_{\alpha\beta}$ is the metric on the 4-sphere).  

Given these facts, we can identify the only non-vanishing Christoffel symbols as 
\beq
\tilde{\Gamma}^{U}_{UU}, \hspace{.25in} \tilde{\Gamma}^{U}_{\mu\nu}, \hspace{.25in} \tilde{\Gamma}^{\mu}_{U\nu}, \hspace{.25in} \tilde{\Gamma}^{U}_{\tau\tau}, \hspace{.25in} \tilde{\Gamma}^{\tau}_{U\tau}, \hspace{.25in} \tilde{\Gamma}^{U}_{\alpha\beta}, \hspace{.25in} \tilde{\Gamma}^{\alpha}_{U\beta}, \hspace{.25in} \tilde{\Gamma}^{\alpha}_{\beta\gamma} \, ,
\eeq
and we can identify that the Ricci tensor must be diagonal.  In addition, the work of \cite{Bigazzi2015} argued that the Kalb-Ramond form vanishes in the background, and the 4-form is unaffected by the backreaction.

We consider graviton modes as fluctuations  around the background metric of equation \eqref{eqn:graviton}. 
We will assume that all of the perturbations depend only on the Lorentz coordinates $x^{\mu}$, and on the radial coordinate $U$.  (That is, we are working with the trivial mode of the graviton KK-tower for the 4-sphere and the $\tau$-direction).  Note that indices of these perturbations will be raised and lowered with the background metric.

We can see that $h^{\tau}_{N}$ can only couple with another $h^{\tau}_{M}$ at quadratic order, as there are no other non-vanishing objects with only one $\tau$ index.  For the same reason, if $N \ne \tau$, we cannot have $M = \tau$.  Finally, we cannot couple $h^{\tau}_{\alpha}$ quadratically to either $h^{\tau}_{\mu}$ or to $h^{\tau}_{U}$, because there are no non-vanishing objects with one $\alpha$ index.  

As a result, we conclude that in the expansion of the bulk action to quadratic order the fields of interest $\{h^{\tau}_{\mu}, h^{\tau}_{U}\}$ decouple from everything else. In particular, this will allow us to ignore the expansions of both the dilaton and the 4-form, and drop any terms involving $h = h^{N}_{N}$.

\subsection{Quadratic-Order Bulk Action for $h_{MN}$}

We now expand the bulk action piece by piece for generic $h_{MN}$. First, we note that to quadratic order we have
\beq
\sqrt{-\det g} = \sqrt{-\det \tilde{g}}\left[1 + \frac{1}{2}h + \frac{1}{8}h^2 - \frac{1}{4}h^{MN}h_{NM}\right] = \sqrt{-\det\tilde{g}}\left[1 - \frac{1}{4}h^{MN}h_{MN}\right] \, ,
\eeq
so that
\beq
\mathcal{L}_{\mathrm{bulk}} = -\frac{1}{4}\sqrt{-\det\tilde{g}} \, e^{-2\tilde{\phi}}\left[\tilde{R} + 4\tilde{g}^{MN}\nabla_M\tilde{\phi}\nabla^M\tilde{\phi} - \frac{e^{2\tilde{\phi}}}{2\cdot 4!}\tilde{F}_4^2\right]h^{MN}h_{MN}
\eeq
$$
 \ + \ \sqrt{-\det\tilde{g}} \, e^{-2\tilde{\phi}} \, g^{MN}\Big[R_{MN} + 4\nabla_{M}\tilde{\phi}\nabla_{N}\tilde{\phi}\Big] \, .
$$

We also have (again to quadratic order)
\beq
g^{MN} = \tilde{g}^{MN} - h^{MN} + h^{MP}h_{P}^{N} \, ,
\eeq
and therefore (dropping linear terms)
\beq
4g^{MN}\nabla_{M}\tilde{\phi}\nabla_{N}\tilde{\phi} = 4\Big(\nabla_{M}\tilde{\phi}\nabla_{N}\tilde{\phi}\Big)h^{MP}h^{N}_{P} \, ,
\eeq
and also
\beq
R = g^{MN}R_{MN} = \Big[\tilde{g}^{MN} - h^{MN} + h^{MP}h_{P}^{N}\Big]R_{MN} = \tilde{R}_{MN}h^{MP}h^{N}_{P} + \Big[\tilde{g}^{MN} - h^{MN}\Big]R_{MN} \, ,
\eeq

which means
\beq
\mathcal{L}_{\mathrm{bulk}} = \sqrt{-\det\tilde{g}} \, e^{-2\tilde{\phi}} \times
\eeq
$$
\Bigg\{\left[-\frac{1}{4}\tilde{R} - \nabla_P\tilde{\phi}\nabla^P\tilde{\phi} + \frac{e^{2\tilde{\phi}}}{8\cdot 4!}\tilde{F}_4^2\right]h^{MN}h_{MN} \ + \ \Big[\tilde{R}_{MN} + 4\nabla_{M}\tilde{\phi}\nabla_{N}\tilde{\phi}\Big]h^{MP}h^{N}_{P}
+ \Big[\tilde{g}^{MN} - h^{MN}\Big]R_{MN}\Bigg\} \, .
$$

What we are left with is the expansion of the Ricci tensor term.

The Cristoffel symbols become
\beq
\Gamma^{M}_{LP} 
= \tilde{\Gamma}^{M}_{LP} + \frac{1}{2}(\nabla_P h^{M}_{L} + \nabla_L h^{M}_{P} - \nabla^M h_{LP}) - \frac{1}{2}h^{MN}(\nabla_{P}h_{NL} + \nabla_{L}h_{NP} - \nabla_{N}h_{LP}) \, ,
\eeq

so the Ricci tensor is

\beq
R_{MN} = \tilde{R}_{NM} + \frac{1}{2}(\nabla_{P}\nabla_{M}h^{P}_{N} + \nabla_{P}\nabla_{N}h^{P}_{M} - \nabla^2h_{NM})
\eeq

$$
- \frac{1}{2}\nabla_P\Big[h^{PQ}\nabla_M h_{QN}\Big] - \frac{1}{2}\nabla_P\Big[h^{PQ} \nabla_N h_{QM}\Big] + \frac{1}{2}\nabla_P\Big[h^{PQ} \nabla_Q h_{NM}\Big]  + \frac{1}{2}\nabla_N\Big[h^{PQ}\nabla_M h_{QP}\Big] 
$$

$$
 - \frac{1}{4}\Big[\nabla_L h^{P}_{N} + \nabla_N h^{P}_{L} - \nabla^P h_{LN}\Big]\Big[\nabla_M h^{L}_{P} + \nabla_P h^{L}_{M} - \nabla^L h_{MP}\Big] \, .
$$

Next we combine the inverse metric and the Ricci tensor, and drop anything zeroeth order or linear order as well as terms involving $h$, to write

\beq
\left[\tilde{g}^{MN} - h^{MN}\right]R_{MN}
\eeq

$$
= -\nabla_{N}\nabla_{M}\Big[h^{MP}h_{P}^{N}\Big] + \frac{1}{2}\nabla^{2}\Big[h^{MN}h_{MN}\Big]
 + \frac{1}{2}\nabla^Ph^{MN}\nabla_{M}h_{PN} - \frac{1}{4}\nabla^Ph^{MN}\nabla_{P}h_{MN} \, .
$$

Putting this together with the earlier work, we now have the quadratic expansion of the bulk Lagrangian density as

\beq
\delta_2\mathcal{L}_{\mathrm{bulk}} = 
\eeq
$$
\sqrt{-\det\tilde{g}} \, e^{-2\tilde{\phi}}\Bigg\{\left[-\frac{1}{4}\tilde{R} - \nabla_P\tilde{\phi}\nabla^P\tilde{\phi} + \frac{e^{2\tilde{\phi}}}{8\cdot 4!}\tilde{F}_4^2\right]h^{MN}h_{MN} \ + \ \Big[\tilde{R}_{MN} + 4\nabla_{M}\tilde{\phi}\nabla_{N}\tilde{\phi}\Big]h^{MP}h^{N}_{P}
$$
$$
-\nabla_{N}\nabla_{M}\Big[h^{MP}h_{P}^{N}\Big] + \frac{1}{2}\nabla^{2}\Big[h^{MN}h_{MN}\Big]
 + \frac{1}{2}\nabla^Ph^{MN}\nabla_{M}h_{PN} - \frac{1}{4}\nabla^Ph^{MN}\nabla_{P}h_{MN}\Bigg\} \, .
$$

Inside the action, we can use integration-by-parts to move total derivatives around, giving us

\beq
\delta_2\mathcal{L}_{\mathrm{bulk}} = \sqrt{-\det\tilde{g}} \, e^{-2\tilde{\phi}}\Bigg\{\frac{1}{2}\nabla_{M}h_{NP}\nabla^{P}h^{MN} - \frac{1}{4}\nabla_{M}h_{NP}\nabla^{M}h^{NP}
\eeq

$$
 + \bigg[\tilde{R}_{MN} + 2\Big(\nabla_M\nabla_N\tilde{\phi}\Big)\bigg]h^{MR}h_{R}^{N}
 -  \bigg[\frac{1}{4}\tilde{R} + \Big(\nabla^2\tilde{\phi}\Big) - \Big(\nabla_{P}\tilde{\phi}\Big)\Big(\nabla^{P}\tilde{\phi}\Big) - \frac{1}{8\cdot 4!}\tilde{F}_4^2e^{2\tilde{\phi}}\bigg]h^{NM}h_{NM}
\Bigg\} \, .
$$

\subsection{Expansion in Terms of $h^{\tau}_{\mu}$ and $h^{\tau}_{U}$}

Finally, we need to identify where the specific terms involving $h^{\tau}_{\mu}$ and $h^{\tau}_{U}$ are.  First we write

\beq
\frac{1}{2}\nabla_{M}h_{NP}\nabla^{P}h^{MN} - \frac{1}{4}\nabla_{M}h_{NP}\nabla^{M}h^{NP}
\eeq

$$
 = -\frac{1}{4}\tilde{g}_{\tau\tau}\tilde{g}^{\mu\rho}\tilde{g}^{\nu\sigma}\Big(\partial_{\rho}h^{\tau}_{\sigma} - \partial_{\sigma}h^{\tau}_{\rho}\Big)\Big(\partial_{\mu}h^{\tau}_{\nu} - \partial_{\nu}h^{\tau}_{\mu}\Big)
 -\frac{1}{2}\tilde{g}_{\tau\tau}\tilde{g}^{\mu\nu}\tilde{g}^{UU}\Big(\partial_{\mu}h^{\tau}_{U} - \partial_{U}h^{\tau}_{\mu}\Big)\Big(\partial_{\nu}h^{\tau}_{U} - \partial_{U}h^{\tau}_{\nu}\Big)
$$
$$
 + \nabla_{M}\Big(\tilde{\Gamma}^{M}_{\tau\tau} \tilde{g}^{\mu\nu}h^{\tau}_{\mu}h^{\tau}_{\nu} + \tilde{\Gamma}^{M}_{\tau\tau}\tilde{g}^{UU}h^{\tau}_{U}h^{\tau}_{U}\Big) - \tilde{R}_{\tau\tau}\Big(\tilde{g}^{\mu\nu}h^{\tau}_{\mu}h^{\tau}_{\nu} + \tilde{g}^{UU}h^{\tau}_{U}h^{\tau}_{U}\Big) \, ,
$$

which then implies

\beq
\sqrt{-\det\tilde{g}} \, e^{-2\tilde{\phi}}\left[\frac{1}{2}\nabla_{M}h_{NP}\nabla^{P}h^{MN} - \frac{1}{4}\nabla_{M}h_{NP}\nabla^{M}h^{NP}\right]
 = e^{-2\tilde{\phi}}\Bigg[-\frac{1}{4}\tilde{g}_{\tau\tau}\tilde{g}^{\mu\rho}\tilde{g}^{\nu\sigma}\Big(\partial_{\rho}h^{\tau}_{\sigma} - \partial_{\sigma}h^{\tau}_{\rho}\Big)\Big(\partial_{\mu}h^{\tau}_{\nu} - \partial_{\nu}h^{\tau}_{\mu}\Big)
 \eeq
 $$
 -\frac{1}{2}\tilde{g}_{\tau\tau}\tilde{g}^{\mu\nu}\tilde{g}^{UU}\Big(\partial_{\mu}h^{\tau}_{U} - \partial_{U}h^{\tau}_{\mu}\Big)\Big(\partial_{\nu}h^{\tau}_{U} - \partial_{U}h^{\tau}_{\nu}\Big)
- \left[\tilde{R}_{\tau\tau} + 2\Big(\nabla_{\tau}\nabla_{\tau}\tilde{\phi}\Big)\right]\Big(\tilde{g}^{\mu\nu}h^{\tau}_{\mu}h^{\tau}_{\nu} + \tilde{g}^{UU}h^{\tau}_{U}h^{\tau}_{U}\Big)\Bigg] \, .
$$

At the same time, we have

\beq
\sqrt{-\det\tilde{g}} \, e^{-2\tilde{\phi}} \, \bigg[\tilde{R}_{MN} + 2\Big(\nabla_M\nabla_N\tilde{\phi}\Big)\bigg]h^{MR}h_{R}^{N}
 -  \frac{1}{2}\bigg[\frac{1}{2}\tilde{R} + 2\Big(\nabla^2\tilde{\phi}\Big) - 2\Big(\nabla_{P}\tilde{\phi}\Big)\Big(\nabla^{P}\tilde{\phi}\Big) - \frac{1}{4\cdot 4!}\tilde{F}_4^2e^{2\tilde{\phi}}\bigg]h^{NM}h_{NM}
\eeq

$$
= \sqrt{-\det\tilde{g}} \, e^{-2\tilde{\phi}}\Bigg\{\bigg[\tilde{R}_{\tau\tau} + 2\Big(\nabla_\tau\nabla_\tau\tilde{\phi}\Big)\bigg]\Big(\tilde{g}^{UU}h^{\tau}_{U}h^{\tau}_{U} + \tilde{g}^{\mu\nu}h^{\tau}_{\mu}h^{\tau}_{\nu}\Big)
$$
$$
 +  \tilde{g}_{\tau\tau}\bigg[\tilde{R}^{UU} + 2\Big(\nabla^U\nabla^U\tilde{\phi}\Big) - \tilde{g}^{UU}\bigg(\frac{1}{2}\tilde{R} + 2\Big(\nabla^2\tilde{\phi}\Big) - 2\Big(\nabla_{P}\tilde{\phi}\Big)\Big(\nabla^{P}\tilde{\phi}\Big) - \frac{1}{4\cdot 4!}\tilde{F}_4^2e^{2\tilde{\phi}}\bigg)\bigg]h^{\tau}_{U}h^{\tau}_{U}
$$
$$
 +  \tilde{g}_{\tau\tau}\bigg[\tilde{R}^{\mu\nu} + 2\Big(\nabla^{\mu}\nabla^{\nu}\tilde{\phi}\Big)-\tilde{g}^{\mu\nu}\bigg(\frac{1}{2}\tilde{R} + 2\Big(\nabla^2\tilde{\phi}\Big) - 2\Big(\nabla_{P}\tilde{\phi}\Big)\Big(\nabla^{P}\tilde{\phi}\Big) - \frac{1}{4\cdot 4!}\tilde{F}_4^2e^{2\tilde{\phi}}\bigg)\bigg]h^{\tau}_{\mu}h^{\tau}_{\nu} \Bigg\} \, .
$$

When we combine these expressions, we obtain equation \eqref{eqn:d2Sbulk}.

\section{
\label{sec:CSappendix} Chern-Simons Term}

While the Chern-Simons terms on the branes contain couplings between the brane and bulk, they do not contribute at quadratic order to the actions of the fields we are interested in. 

A crucial element of this analysis is the fact that both the first and second Pontryagin classes vanish on both the original and smeared backreacted backgrounds due to the diagonal structure of the metric and the fact that the smeared background retains the same isometries as the original. We demonstrate this explicitly in section 5.1. 

The Chern-Simons term takes the form \cite{Green:1996dd}:
 \begin{align}\label{CS-general}
 S_{\text{CS}}=\int C \wedge \Tr e^{F/2\pi} \wedge \sqrt{\hat{\mathcal{A}}(R)}~.
 \end{align}
Here $C = \sum_{i} C_i$ is the sum of bulk Ramond-Ramond forms, $F$ is the gauge field strength on the branes, and the A-roof genus $\hat{\mathcal{A}}$ is given by
 \begin{align}
 \hat{\mathcal{A}} (R) = 1 - \frac{p_1}{24}  + \frac{1}{16}\left(\frac{7}{360} p_1^2 -\frac{1}{90}p_2\right)+\dots
 \end{align}
 where $p_1$ and $p_2$ are the Pontryagin classes defined as:
 \begin{align}
 p_1(R) &= -\frac{1}{8\pi}\Tr R\wedge R \cr
 p_2(R) &= \frac{1}{128\pi^4}\left[ (\Tr R\wedge R)^2 - 2 \Tr R\wedge R\wedge R\wedge R\right]~ 
 \end{align}
 with $R$, the Riemann tensor 2-form, 
 \begin{align}
 R_{AB} = \frac{1}{2} R_{ABCD} dx^C\wedge dx^D~.
 \end{align}
Expanding \eqref{CS-general}, we can identify terms which might contain the brane scalar $\Phi$ or the vector piece of the graviton $(h^\tau_U,~h^\tau_\mu)$, together or with another field:
\begin{align}\label{CSexp}
S_{\text{CS}} 
&\supset \int\bigg\{ -\frac{1}{3840}C_1 \wedge  \left( p_1^2 +\frac{4}{3}p_2 \right) -\frac{1}{48} C_3 \wedge \Tr \frac{F}{2\pi} \wedge p_1 -\frac{1}{48} C_5\wedge p_1 \bigg\}\cr
\end{align}
Note that the bulk fields appearing in \eqref{CS-general} are pulled back to the brane's worldvolume, with possible couplings to $\Phi$ would come via the pullback as, for instance,
\begin{align}
P[C_1]_A = C_B \frac{\d X^B}{\d x^A} = C_A + \frac{1}{2\pi\alpha'}C_\tau \frac{\d\Phi}{\d x^A}
\end{align}
where $X^A$ are fluctuation scalars and $x^A$ are the brane coordinates. (In the second equality we applied static gauge.) 

 One can show that in the background with the isometries of WSS, neither $p_1$ nor $p_2$ have non-trivial background values, nor do they contain terms that are first order in $(h^\tau_\mu, h^\tau_U)$ (and zeroth order in other fields). Note that this is true for both the original WSS model, which treats the D8's as probes, and for the partially-backreacted, ``smeared" geometry used here. 
 
 \subsection{Relevant properties of the Pontryagin classes}

Our argument for the vanishing of the Chern-Simons contribution relies on the fact that neither of the Pontryagin classes $p_1$ and $p_2$ have non-zero background value, nor do they have contributions at first order in $(h^\tau_\mu,~h^\tau_U)$. It will also rely on the components of $p_1$ along the $S^4$ having no quadratic-order contributions.

We now demonstrate each of these facts individually, before arguing for the vanishing of the whole quadratic-order Chern-Simons contribution in the next subsection.

Note that the metric we are working with -- both in its original form, and in the form that includes some backreaction -- is diagonal and depends only on the radial coordinate, $U$ (and the coordinates of the $S^4$ to the extent that they appear in the $S^4$ metric). It's then straightforward if a bit tedious to show that the only non-vanishing Riemann tensor components are:
\begin{align}
    \tilde{R}^\mu_{\phantom{\mu}\nu\rho\sigma},~ \tilde{R}^\mu_{\phantom{\mu}U\nu U},~ \tilde{R}^\mu_{\phantom{\mu}\tau\nu\tau},~ \tilde{R}^\mu_{\phantom{\mu}\alpha\nu\beta},~\cr \tilde{R}^U_{\phantom{\mu}\tau U\tau}, \tilde{R}^U_{\phantom{\mu}\alpha U\beta},\cr
    \tilde{R}^\tau_{\phantom{\mu}\alpha \tau\beta},~ 
     \tilde{R}^\tau_{\phantom{\mu}\alpha \tau\beta},~ \tilde{R}^\alpha_{\phantom{\mu}\beta \gamma\delta}~,
\end{align}
plus those related to these by the symmetries of the Riemann tensor,
\begin{align}
    R_{MNPQ}=g_{ML}R^L_{\phantom{L}NPQ}=-R_{NMPQ}=-R_{MNQP}=R_{PQMN}~.
\end{align}
As before we indicate with a tilde that these are quantities evaluated on the supergravity background. We can heuristically summarize the above with
\begin{align}
    \tilde{R}^M_{\phantom{M}NLP} \propto (\delta^M_L\delta_{NP}-\delta^M_P\delta_{NL}) ~.
\end{align}
Structures appearing in $p_1$ and $p_2$ are $\Tr (R\wedge R)$ and $\Tr (R\wedge R\wedge R\wedge R)$. We can now see that these vanish on the background. The object
\begin{align}
\{ (\tilde{R}^A_{\phantom{A}B})\wedge (\tilde{R}^B_{\phantom{A}C}) \}_{MNLP} = \tilde{R}^A_{\phantom{A}B[MN}\tilde{R}^B_{\phantom{A}|C|LP]} \propto \delta^A_{[M}\delta_{NL}\delta_{|C|P]}
\end{align}
appears once in $\Tr (R\wedge R)$ and twice in $\Tr (R\wedge R\wedge R\wedge R)$, and clearly vanishes. Hence, 
\begin{align}
    p_1=\mathcal{O}(h)~, \quad p_2=\mathcal{O}(h^2)~
\end{align}
where $h$ represents any component of the graviton $h_{MN}$.

The next step is to examine whether $p_1$, pulled back to the branes, depends on $(h^\tau_\mu,~h^\tau_U)$ at linear order. They do not.
The Christoffel symbols that are first order in $h=(h^\tau_\mu,~h^\tau_U)$ are
\beq
\Gamma^{\mu}_{\nu\tau} \, , \hspace{.5in} \Gamma^{\mu}_{U\tau} \, , \hspace{.5in} \Gamma^{U}_{\mu\tau} \, , \hspace{.5in} \Gamma^{U}_{U\tau}
\eeq
\beq
\Gamma^{\tau}_{\mu\nu} \, , \hspace{.5in} \Gamma^{\tau}_{\mu U} \, , \hspace{.5in} \Gamma^{\tau}_{UU} \, , \hspace{.5in} \Gamma^{\tau}_{\tau\tau}
\eeq
\beq
\Gamma^{\tau}_{\alpha\beta} \, , \hspace{.5in} \Gamma^{\alpha}_{\tau\beta}~.
\eeq
None of these have $\mathcal{O}(h^0)$ terms. Clearly, all Riemann tensor components that are first order in $(h^\tau_\mu,~h^\tau_U)$ must have a $\tau$ index as we can see from the definition of the Riemann tensor
\beq
R^{M}_{NLP} = \partial_{L}\Gamma^{M}_{NP} - \partial_{P}\Gamma^{M}_{NL} + \Gamma^{M}_{LQ}\Gamma^{Q}_{NP} - \Gamma^{M}_{PQ}\Gamma^{Q}_{NL}~.
\eeq
Terms in $p_1$ that are linear in $h$ must be of the form
\begin{align}
   \tilde{R}^A_{\phantom{A}BMN}\hat{R}^B_{\phantom{A}ALP}
\end{align}
where the $\hat{R}^B_{\phantom{A}ALP}$ is first order $(h^\tau_\mu,~h^\tau_U)$. Clearly one of $A,~B,~L,~P$ must equal $\tau$ for this quantity to be finite. If $L=\tau$ or $P=\tau$, this component of $p_1$ would need to be pulled back to the brane worldvolume using a $\Phi$ field, which places it at second order. If $A=\tau$ or $B=\tau$,  $\tilde{R}^A_{\phantom{M}BMN}$ would need to have $M=\tau$ or $N=\tau$, which again require a pullback. Thus $p_1$ has no first order components in the fields of interest.

Finally, we show that $(p_1)_{\alpha\beta\gamma\delta}\sim \mathcal{O}((h^\tau_\mu,~h^\tau_U)^3)$~. 
Since $p_1\propto \Tr R\wedge R$, we need to examine 
\beq
R^{M}_{\phantom{M}N\alpha\beta}R^{N}_{\phantom{M}M\gamma\delta}~
\eeq
for quadratic-order terms -- either as a linear terms from each  Riemann tensor, or quadratic terms from one (with the other taking on its background value). 

First we look for first order terms in $R^{M}_{\phantom{M}N\alpha\beta}$ which takes the form
\beq
R^{M}_{\phantom{M}N\alpha\beta} = \partial_{\alpha}\Gamma^{M}_{N\beta} + \Gamma^{M}_{\alpha L}\Gamma^{L}_{N\beta} - (\alpha \leftrightarrow \beta)
\eeq
where
\beq
\Gamma^{M}_{\alpha L} = \tilde{\Gamma}^{M}_{\alpha L} - h^{M}_{U}\tilde{\Gamma}^{U}_{\alpha L} + \cdots~.
\eeq
Thus
\beq
R^{M}_{N\alpha\beta} = \tilde{R}^{M}_{N\alpha\beta} - \Big(h^{M}_{U}\partial_{\alpha}\tilde{\Gamma}^{U}_{\beta N} + h^{L}_{U}\tilde{\Gamma}^{M}_{\alpha L}\tilde{\Gamma}^{U}_{N\beta} + h^{M}_{U}\tilde{\Gamma}^{U}_{\alpha L}\tilde{\Gamma}^{L}_{N\beta}\Big) + (\alpha \leftrightarrow \beta) + \cdots
\eeq
\beq
= \tilde{R}^{M}_{N\alpha\beta} - h^{M}_{U}\tilde{R}^{U}_{N\alpha\beta} + \cdots~,
\eeq
which vanishes because the only non-zero components of the background Riemann tensor on the $S^4$ are $\tilde{R}^\alpha_{\phantom{a}\beta\gamma\delta}$.

This fact also implies that we need only concern ourselves with possible second-order contributions from $R^\alpha_{\phantom{a}\beta\gamma\delta}$. One can show that the expansion of these components takes the form 
\beq
R^{\alpha}_{\beta\gamma\delta} = \tilde{R}^{\alpha}_{\beta\gamma\delta} - h^{\tau}_{U}h^{\alpha}_{\tau}\tilde{R}^{U}_{\beta\gamma\delta} - h^{N\tau}h_{\tau M}\Big(\tilde{\Gamma}^{M}_{\beta\delta}\tilde{\Gamma}^{\alpha}_{\gamma N} - (\gamma \leftrightarrow \delta)\Big)
\eeq
\beq
= \tilde{R}^{\alpha}_{\beta\gamma\delta} - h^{N\tau}h_{\tau M}\Big(\tilde{\Gamma}^{M}_{\beta\delta}\tilde{\Gamma}^{\alpha}_{\gamma N} - (\gamma \leftrightarrow \delta)\Big)
\eeq
We must have $N = U$, or $M = U$, or both for the second term to be finite.  Any term of that kind will also (by virtue of the structure of the background Christoffel symbols) force $\alpha = \gamma$, or $\alpha = \delta$, or $\beta = \gamma$, or $\beta = \delta$.  When we contract this with the background Riemann tensor antisymmetrize on the form indices, this contribution will also vanish.

\subsection{Vanishing Contribution From The Chern-Simons Term}

We can now check one by one that the terms in \eqref{CSexp} do not contribute to the quadratic-order action for $\Phi$ and $(h^\tau_\mu,~ h^\tau_U)$, using the properties derived in the previous subsection.

The first term goes like
\begin{align}
    \int c_1 \wedge \left( p_1^2 + \frac{4}{3} p_2\right)~.
\end{align}
The supergravity background does not have a background $C_1$, so the $c_1$ appearing here is a fluctuation (representing a glueball state). Neither $p_1^2$ nor $p_2$ have first order contributions in $(h^\tau_\mu,~h^\tau_U)$, so this term does not mix $c_1$ with them. $p_1$ and $p_2$ also have no background values, so this term also cannot mix $c_1$ with $\Phi$ via the pullback.

The second term in \eqref{CSexp} goes like 
\begin{align}
    \int C_3 \wedge F \wedge p_1 = \int \left( \tilde{C}_3 \wedge F \wedge p_1 + c_3 \wedge F \wedge p_1 \right)
\end{align}
where as above we use $\tilde{C}_3$ to denote the background value and $c_3$ to denote the fluctuation of the potential corresponding to a glueball mode. The first term in this expression is at least first order in fluctations (because of the gauge field strength $F$), and cannot mix the brane gauge field with $(h^\tau_\mu, ~h^\tau_U)$ at quadratic order because $p_1$ has no first-order terms. It does not mix the gauge field with $\Phi$, meanwhile, because $p_1$ has vanishing background value. The second term in the $C_3$ term expression is already second order in fields (and vanishes anyway because $p_1$ vanishes on the background). 

Finally, we have the third term in \eqref{CSexp}, of the form
\begin{align}
    \int C_5\wedge p_1 = \int (\tilde{C}_5 + c_5)\wedge p_1~. 
\end{align}
The second term in this expression does not contribute any quadratic terms to the action because $p_1$ has no first order terms. This also means that the first term, containing the background value of $C_5$, does not provide any terms mixing $\Phi$ with $(h^\tau_\mu,~h^\tau_U)$ via the pullback. 

One could still have terms at quadratic order in $h$ in $p_1$. It turns out that the relevant components of $p_1$ do not contain second order terms involving $(h^\tau_\mu,~h^\tau_U)$. The background value $\tilde{C_5}$ is defined via the field strength
$\tilde{F}_6 = dC_5$ which is related to $\tilde{F}_4$ via the Hodge dual: $\tilde{F}_6=\star \tilde{F}_4$. The background $\tilde{F}_4$ is proportional to the volume form on the $S^4$ and otherwise only depends on $U$. $\tilde{F}_6$ thus has components on the remaining 6 directions, $(x^\mu,~U,~\tau)$, and so we can choose $\tilde{C}_5$ to components in 5 of these. If one component of $\tilde{C}_5$ lies along the $\tau$ direction, it would need to be pulled back to the brane world volume via a $\Phi$. Such a term would not contribute to a quadratic term because $p_1$ is second order. If $\tilde{C}_5$ has components $(x^\mu,~U)$, one could get a contribution at quadratic order from a component of $p_1$ with all components along the $S^4$. However, as shown above, none of these contain  $(h^\tau_\mu,~h^\tau_U)$.

The Chern-Simons term thus contributes no additional quadratic order terms in the action of $\Phi$ and $(h^\tau_\mu,~h^\tau_U)$, and we ignore it in what follows.

\section{
\label{sec:NUMappendix} Numerical Methods}

Here we present a brief explanation of the numerical techniques used to acquire the results given in table \ref{vectortable}.  To find the unperturned eigenstatates and eigenvalues for both the vector and scalar wavefunctions, we used the ``shooting method''.  Then, we used numerical integration to compute overlap integrals necessary to find the perturbations to the eigenvalues.

\begin{itemize}

\item Shooting Method Details for the Scalar Wavefunctions:

\begin{itemize}
\item Shooting for the scalar wavefunctions $\varphi_i(u)$ was performed first in the $z = \frac{Z}{U_{\mathrm{KK}}}$ variable, normalized, and then the solutions were converted to $u$.
\item The built-in differential equation solver \texttt{NDSolve} was used within \texttt{Mathematica} to find the solutions, and \texttt{NIntegrate} was used to normalize them.
\item Correctly implementing the boundary condition at $u = 1$ then gave us wavefunctions which vanished at this location, as can be seen in figure \ref{scalarpic}.
\item In order to implement ``shooting'' to find the eigenvalues $\chi_i^{(0)}$, we introduced two parameters $u_{\infty}$ and $\varepsilon$, and required $|\varphi_i(u_{\infty})| \le \varepsilon$.
\end{itemize}

\item Shooting Method Details for the Vector Wavefunctions:

\begin{itemize}
\item Shooting for the vector wavefunctions $\psi_n(u)$ was performed in the $u$ variable, again using \texttt{NDSolve} inside \texttt{Mathematica}, and then normalized using \texttt{NIntegrate}.
\item The differential equation  is badly behaved at $u = 1$, so the method utilized a cut-off parameter $u_0 = 1 + \delta$, and implemented boundary conditions at this location based on a series expansion of the solutions.
\item Again, correctly implementing the boundary conditions gave us wavefunctions which vanished at $u = 1$ (as seen in figure \ref{vectorpic}).
\item The eigenvalues $\lambda_n^{(0)}$ were found using ``shooting'' with the same parameters $u_{\infty}$ and $\varepsilon$, and the requirement $|\psi_n(u_{\infty})| \le \varepsilon$.
\end{itemize}

\item Calculating $\delta\lambda_n$ and $\delta\chi_{i,1}$:

\begin{itemize}
\item To compute the overlap integrals giving the corrections to each eigenvalue from the background geometry perturbation, the functions $a^{(1)}(u)$, $b^{(1)}(u)$, and $c^{(1)}(u)$ were entered into \texttt{Mathematica} analytically.
\item However, the behavior of these functions for large $u$ was found to be more accurate and stable if, above a parameter $u_{\mathrm{max}}$, a Taylor Series approximation was used instead.
\item This Taylor series was implemented with $N_s$ terms in it (with $N_s$ a new numerical parameter).
\item Having done this, \texttt{NIntegrate} was used to perform the necessary overlap integrals, with the limits $u \in [1 + \delta, u_{\infty}]$, fixed by the numerical parameters chosen earlier.
\end{itemize}

\item Calculating $\delta\chi_{i,2}$:

\begin{itemize}
\item To compute $\delta\chi_{i,2}$, we approximated \eqref{eqn:dHvarphi2} as
\beq
\delta\hat{H}_{\check{\varphi}, 2} = \sum_{n}^{N} \hat{\mathcal{K}}^{(0)}\left(\frac{|n\rangle_{0} \, {}_{0}\langle n|}{\lambda_n^{(0)}}\right)\hat{\mathcal{K}}^{(0)}\hat{H}_{\varphi}^{(0)}
\eeq
\item This introduced a final numerical parameter $N$, corresponding to the number of vector states summed over to approximate $\Big(\hat{H}_{\psi}^{(0)}\Big)^{-1}$.
\end{itemize}

\item Numerical Parameter Choices

\begin{itemize}
\item The choices for the numerical parameters utilized are listed in table \ref{numtable}.
\item The results were tested for robustness by varying these parameters (singly and in combinations), and determining how much the results varied.
\item The limiting factor in accuracy was found to be the value of $N$, the number of vector states included in the calculation of $\delta\chi_{i,2}$.
\item Numerical integrals were then used, as usual, with \texttt{NIntegrate}, using the limits $u \in [1+\delta, u_{\infty}]$.
\end{itemize}

\end{itemize}

\begin{table}
\begin{center}
\begin{tabular}{|c|c|}
\hline
parameter & value \\
\hline
\hline
$u_{\infty}$ & 1000000 \\
\hline
$\varepsilon$ & $10^{-20}$ \\
\hline
$\delta$ & $10^{-10}$ \\
\hline
$u_{\mathrm{max}}$ & 10 \\
\hline
$N_s$ & 20 \\
\hline
$N$ & 50 \\
\hline
\end{tabular}
\caption{\label{numtable} A table of numerical parameters used.}
\end{center}
\end{table}

\bibliographystyle{unsrt}
\bibliography{refs}

\end{document}